%% file: main_arxiv.tex
\documentclass{article} 
\usepackage{iclr2027_conference,times,booktabs}
\input{preamble.tex}  
\usepackage[table]{xcolor}
\usepackage[most]{tcolorbox} 
\definecolor{fbback}{HTML}{F8FBF5}
\definecolor{fbframe}{HTML}{CBDDBE}
\definecolor{fbtitle}{HTML}{E4EEDC}   
\definecolor{fbtext}{HTML}{3F5F36}    
\definecolor{lightgray}{HTML}{EFEFEF}
\tcbset{cbox/.style={
  enhanced,
  colback=fbback,
  colframe=fbframe,
  colbacktitle=fbtitle,
  coltitle=fbtext,
  fonttitle=\bfseries\small,
  fontupper=\small\ttfamily,
  boxrule=0.6pt,
  arc=2pt,
  left=6pt, right=6pt, top=5pt, bottom=5pt,
}}
\usepackage{longtable} 
\usepackage{caption}  
\tcbuselibrary{listings,breakable,skins} 
\usepackage{amssymb}

\input{math_commands.tex}

\usepackage{hyperref}
\usepackage{url}
\usepackage{graphicx}
\usepackage{algorithm}
\usepackage{algpseudocode}
\usepackage{array}
\usepackage{tabularx}
\usepackage{multirow}

\usepackage{graphicx}   
\usepackage{subcaption} 

\title{
\raisebox{-0.06\height}{
\includegraphics[height=0.68cm]{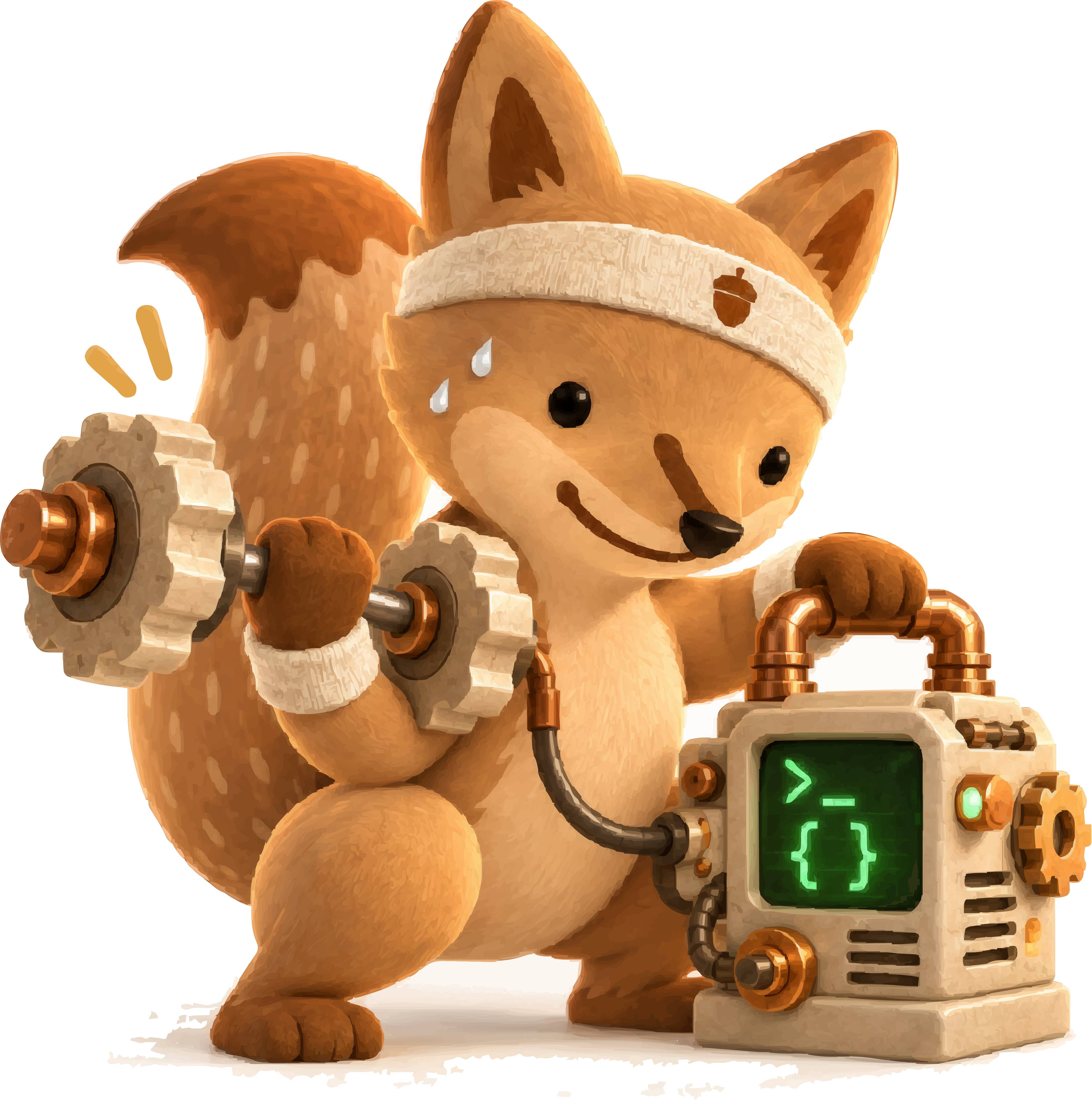}
}
\hspace{-0.8cm}
ProgramGym: Scaling Training Environments for Long-Horizon Program Reproduction Agents
}

\author{Chengyi~Yang$^{1}$\thanks{Equal Contributions.}\;,
Renxu~Bao$^{1}$\footnotemark[1]\;,
Jue~Hou$^{1}$\footnotemark[1]\;,
Zongpei~Teng$^{1}$, 
Ziheng~Wang$^{1}$ \\
\textbf{Yujie~Lin$^{1}$}, 
\textbf{Chengsong~Huang$^{2}$},
\textbf{Junnan~Zhu$^{3}$}, 
\textbf{Delai~Qiu$^{4}$} \\
\textbf{Yining~Wang$^{4}$}, 
\textbf{Shengping~Liu$^{4}$}, 
\textbf{Fei~Long$^{1}$}\thanks{Corresponding authors.}\;, 
\textbf{Jinsong~Su$^{1}$}\footnotemark[2] \\
$^{1}$Xiamen University \quad
$^{2}$Washington University in St.\ Louis \\
$^{3}$Institute of Automation, Chinese Academy of Sciences \quad
$^{4}$Unisound \\
\texttt{yangchengyi@stu.xmu.edu.cn}\quad\texttt{\{flong,jssu\}@xmu.edu.cn}
}

\iclrfinalcopy 
\begin{document}

\maketitle
\lhead{Preprint}
\input{section/0_abstract}

\input{section/1_intro}
\input{section/7_related_work}

\input{section/2_preliminary}
\input{section/3_method}
\input{section/4_experiment}
\input{section/5_discussion}
\input{section/6_conclusion}
\input{extra}

\bibliography{iclr2027_conference}
\bibliographystyle{iclr2027_conference}

\clearpage
\appendix
\input{section/8_Appendix}

\end{document}

%% file: preamble.tex
\usepackage{booktabs,array,graphicx}
\usepackage{hyperref}

\providecommand{\PBIconDir}{icons}
\newcommand{\PBIconWidth}{3em}
\newcommand{\PBIconHeight}{3em}
\newcommand{\PBCode}[1]{\texttt{\detokenize{#1}}}
\newcommand{\PBLanguage}[4]{%
  \begingroup
  \def\PBThisWidth{\PBIconWidth}%
  \def\PBThisIcon{#1}\def\PBGoIcon{go}%
  \ifx\PBThisIcon\PBGoIcon\def\PBThisWidth{3.8em}\fi
  \shortstack[c]{%
    \IfFileExists{\PBIconDir/#1.pdf}{%
      \includegraphics[width=\PBThisWidth,height=\PBIconHeight,keepaspectratio]{\PBIconDir/#1.pdf}%
    }{\IfFileExists{\PBIconDir/#1.png}{%
      \includegraphics[width=\PBThisWidth,height=\PBIconHeight,keepaspectratio]{\PBIconDir/#1.png}%
    }{\begingroup\setlength{\fboxsep}{2pt}%
      \fbox{\parbox[c][2.4em][c]{2.4em}{\centering\bfseries #2}}\endgroup}}%
    \\[.25em]%
    \href{#4}{\footnotesize #3}%
  }%
  \endgroup
}
\newcommand{\PBGCCDoc}{https://gcc.gnu.org/onlinedocs/gcc/Instrumentation-Options.html}

\newcommand{\PBRustDoc}{https://doc.rust-lang.org/rustc/instrument-coverage.html}
\newcommand{\PBGoDoc}{https://go.dev/doc/build-cover}

\usepackage{svg}
\renewcommand{\PBIconDir}{svg-icon}
\renewcommand{\PBLanguage}[4]{%
  \begingroup
  \def\PBThisWidth{3em}%
  \def\PBThisIcon{#1}\def\PBGoIcon{go}%
  \ifx\PBThisIcon\PBGoIcon\def\PBThisWidth{3.8em}\fi
  \shortstack[c]{%
    \IfFileExists{\PBIconDir/#1.pdf}{%
      \includegraphics[width=\PBThisWidth,height=3em,keepaspectratio]{\PBIconDir/#1.pdf}%
    }{\IfFileExists{\PBIconDir/#1.svg}{%
      \includesvg[inkscapelatex=false,width=\PBThisWidth,height=3em,keepaspectratio]{\PBIconDir/#1}%
    }{\IfFileExists{\PBIconDir/#1.png}{%
      \includegraphics[width=\PBThisWidth,height=3em,keepaspectratio]{\PBIconDir/#1.png}%
    }{\fbox{\rule{0pt}{2.6em}\rule{2.6em}{0pt}}}}}%
    \\[.25em]%
    \href{#4}{\footnotesize #3}%
  }%
  \endgroup
}
\let\PBOrigIncludegraphics\includegraphics
\newcommand{\PBSmallIconHeight}{1.8em}
\newcommand{\PBLanguageSmall}[3]{%
  \begingroup
  \def\href##1##2{##2}%
  \renewcommand{\includegraphics}[2][]{%
    \PBOrigIncludegraphics[##1,height=\PBSmallIconHeight,keepaspectratio]{##2}}%
  \PBLanguage{#1}{#2}{#3}{}%
  \endgroup}

\definecolor{fbback}{HTML}{EFF7F8}   
\definecolor{fbframe}{HTML}{9FC4C9}  
\definecolor{fbtitle}{HTML}{3E7C86}  

%% file: math_commands.tex
\usepackage{amsmath,amsfonts,bm}

\def\eqref#1{equation~\ref{#1}}

\def\1{\bm{1}}

\DeclareMathAlphabet{\mathsfit}{\encodingdefault}{\sfdefault}{m}{sl}
\SetMathAlphabet{\mathsfit}{bold}{\encodingdefault}{\sfdefault}{bx}{n}



%% file: section/0_abstract.tex
\begin{abstract}
Coding agents are increasingly used to build programs from scratch as software engineers do rather than to resolve issues within a repository or to generate a repository from a given requirements document. However, scalable verifiable training environments that force agents to discover what is required and to decide how the system should be designed remain scarce. Moreover, existing agent workflows also cannot sustain long-horizon development in which the agent explores the reference program before implementation. To address these challenges, we introduce \textbf{ProgramGym}, a framework consisting of Program-Env and Program-Agent for scalable training environments of long-horizon program reproduction agents. Specifically, Program-Env constructs verifiable training environments from repositories through a feedback-driven loop that repairs runtime sandbox builds and the quality of black-box tests. Using this novel pipeline, we obtain $1{,}648$ training environments spanning C, C++, Go, and Rust. Program-Agent combines a role-isolated workflow of analysis, implementation, and review with revision management for compilation checks and failure recovery. Experiments on ProgramBench show that both Program-Env and Program-Agent improve performance on program reproduction. Fine-tuning on trajectories with these constructed environments raises the performance of \texttt{Qwen3.8-27B} from $34.87\%$ to $42.84\%$. Program-Agent further raises the base and SFT models to $55.71\%$ and $56.25\%$, both exceeding some flagship models. We will release the ProgramGym to support further research on coding agents that build programs from scratch.
\end{abstract}

\vspace{-15pt}
{\centering\normalsize
\raisebox{-0.28\height}{\includegraphics[height=1.5em]{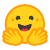}}~%
\href{https://huggingface.co/ProgramGym}{\textcolor{blue}{\texttt{huggingface.co/ProgramGym}}}%
\hspace{2.00em}%
\raisebox{-0.28\height}{\includegraphics[height=2.0em]{svg-icon/programGym.pdf}}~%
\href{https://programgym.github.io/}{\textcolor{blue}{\texttt{programgym.github.io/}}}\par}
\vspace{10pt}
\begin{figure}[h] 
    \centering
    
    \includegraphics[width=1.0\linewidth]{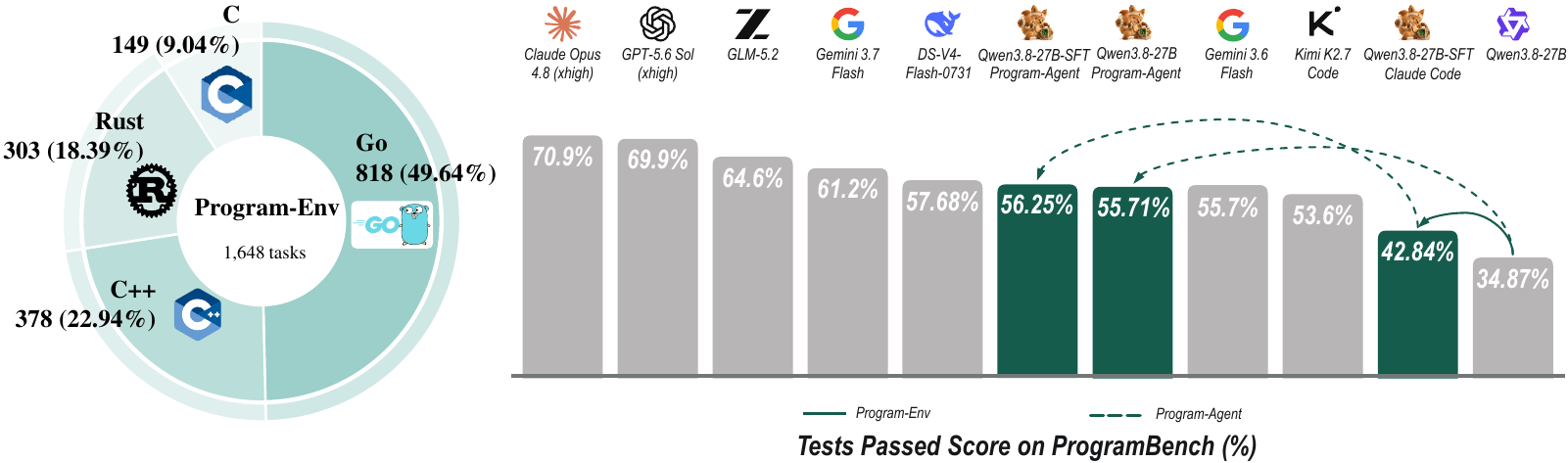} 
    \caption{Overview of \textbf{ProgramGym}.}
    \label{fig:ProgramGym }
\end{figure}

%% file: section/1_intro.tex
\section{Introduction}


Current coding agents~\citep{qwen38,2026kimi-k3,zeng2026glm} typically handle issue-resolution tasks and repository generation tasks. Issue-resolution tasks such as SWE-bench give a repository and an issue for the agent to patch~\citep{jimenez2024swe,rashid2025swe,zeng2026davinci,shi2026swebenchpromaxbenchmarkingagents}. Repository generation tasks such as NL2Repo-Bench give a requirements document for the agent to implement~\citep{ding2025nl2repo}. In both settings, the agent starts from a given codebase or a specified feature. A software engineer building a program from scratch instead keeps asking what is required and how the system should be designed.


To explore this from-scratch setting, program reproduction tasks such as ProgramBench ask an agent to rebuild a program's observable behavior from its usage documentation and a reference program  without access to the source code or network~\citep{yang2026programbench,adamczewski2026mirrorcode}. The agent needs to still discover remaining behavior by exploration and design the implementation from scratch. None of the models in the original ProgramBench evaluation fully solved a task~\citep{yang2026programbench}\footnote{Release-time results; later performance of the flagship models are at \url{https://programbench.com}.}. Training therefore requires interactive environments with reliable behavioral supervision, which issue-resolution resources such as SWE-Gym and SWE-smith do not provide~\citep{pan2024training,yang2026swe,song2026swe,chen2026sweuniverse}. MindForge~\citep{chen2026mindforgeteachingsmalllanguage}is restricted to command-line programs and still leaves this supervision unverifiable. Building verifiable environments and using them over long horizons raises two challenges. \textbf{(i) Scaling verifiable training environments.} Building the reference executable and runtime sandbox requires accommodating repository-specific toolchains and dependencies. The black-box tests need to cover diverse functionality while rejecting incorrect reproductions, yet a single test-generation attempt may miss important behaviors or produce weak assertions. \textbf{(ii) Sustaining long-horizon development.} General-purpose agent harnesses support tool use but do not by themselves keep exploration and implementation coordinated. An agent can begin implementing before sufficiently exploring the reference program and forget previously observed requirements as its context grows. Addressing these risks requires an explicit development workflow that preserves behavioral evidence and coordinates implementation, review, and revision.


 


To overcome these challenges, we introduce \textbf{ProgramGym} consisting of Program-Env and Program-Agent for scalable training environments of long-horizon program reproduction agents. Specifically, Program-Env addresses environment construction by augmenting an existing agent harness with feedback-driven loops for runtime sandbox and black-box test construction. To produce the reference executable and its runtime sandbox, it separates online dependency preparation from offline compilation and returns build diagnostics to guide repairs. Test construction proceeds across functional modules in parallel, using program analysis to target unexplored behavior and validation feedback. The resulting environments provide reference executables, runtime sandboxes, documentation, and black-box behavioral test suites for training supervision. Program-Agent addresses long-horizon development through a role-isolated iterative workflow of analysis, implementation, and review. It first records evidence from reference-program exploration and then alternates implementation with behavioral review in separate sessions. Persistent records carry requirements across rounds, while offline build checks and validated checkpoints support revision and failure recovery. 

Building on these designs, Program-Env scales $1{,}648$ training environments covering C, C++, Go, and Rust. Each environment provides a reference executable, a runtime sandbox, usage documentation, and black-box tests, which are essential for training agents on program reproduction from scratch. We conduct extensive experiments on ProgramBench. With Program-Agent, \texttt{Qwen3.8-27B} reaches $55.71\%$ Tests Passed and \texttt{Qwen3.8-27B-SFT} reaches $56.25\%$, both exceeding flagship \texttt{Claude Opus 4.7}. We further evaluate \texttt{DeepSeek-V4-Flash-0731} under the same protocol. These results suggest that ProgramGym can scale training environments for coding agents that reproduce programs from scratch as software engineers do. We will release ProgramGym as a scalable gym for the community to develop coding agents that discover requirements and design systems from scratch. In summary, our main contributions are threefold:
\begin{itemize}
    \item We propose \textbf{Program-Env}, a feedback-driven pipeline that builds the reference executable and runtime sandbox and constructs black-box tests from repositories (Section~\ref{sec:program-env}).
    \item We propose \textbf{Program-Agent}, which combines a role-isolated workflow of analysis, implementation, and review with revision management for compilation checks and failure recovery (Section~\ref{sec:program-agent}).
    \item We conduct extensive experiments on ProgramBench and show that Program-Env training and Program-Agent both improve program reproduction. We further analyze environment construction as well as iteration budget and agent behavior (Section~\ref{sec:experiments}).
\end{itemize}

%% file: section/7_related_work.tex
\section{Related Work}


\paragraph{Long-Horizon Program Reproduction.} Training large language models as agents has gained substantial attention~\citep{2026kimi-k3,qwen2025qwen25technicalreport,deepseek-coder}, with software engineering (SWE) emerging as a promising application domain~\citep{cao2026qwen3,openai, pan2024training}. Issue resolution is a representative setting, in which SWE-bench tasks~\citep{jimenez2024swe} and its variants evaluate coding-agents on addressing bug fixes or feature requests within a given code base~\citep{yang2026swe,shi2026swebenchpromaxbenchmarkingagents,ahmad2026open,wu2026swe,la2026game,yang2024swe}. Despite their growing diversity, issue resolution tasks depend on a provided code base, while several prior and concurrent works go further by asking agents to code from scratch according to a requirements document~\citep{luo2026gamecraft,zhao2026denovoswe,zhao2025commit0,li2024deveval}. However, this setting typically relies on a natural-language requirements document that prescribes the functional specification~\citep{ding2025nl2repo}, reducing the task to implementation and leaving no room for requirement discovery or architectural design. In contrast, recent work on {program reproduction from scratch} requires agents to reconstruct the externally observable behavior of a reference program from a binary executable and documentation without prescribing implementation structure~\citep{chen2026mindforgeteachingsmalllanguage,yang2026programbench,adamczewski2026mirrorcode}. This setting therefore shifts the challenge from implementing a specified repository to iteratively discovering what the program should do and deciding how to rebuild it, as experienced software engineers.

\paragraph{Code Agent Environments Scaling.}

Existing program reproduction from scratch remains highly challenging~\citep{yang2026programbench,adamczewski2026mirrorcode}. None of the flagship models in the original ProgramBench evaluation fully resolved any of the 200 tasks~\citep{yang2026programbench}. Due to the scarcity of training environments for such tasks, coding agents tend to be trained as code-edit assistants that favor localized bug fixes or single-file implementations with detailed human guidance rather than as autonomous software engineers~\citep{adamczewski2026mirrorcode}. Providing realistic and executable training environments at scale is essential. This strategy is already proven in issue resolution~\citep{song2026swe,zhao2026immersion,chen2026sweuniverse}, where generated environments improved SWE-agent performance~\citep{jain2025r2e,pan2024training}. These frameworks still start from an existing codebase or a well-specified task description. MindForge does not filter training trajectories by recovered program behavior, so functionally incorrect reproductions can still be treated as successful supervision~\citep{chen2026mindforgeteachingsmalllanguage}. By contrast, ProgramGym makes program reproduction from scratch training scalable by automatically turning repositories into verifiable training environments. This increases sample sizes and converts coding agents from compliant code editors into active software engineers.



%% file: section/2_preliminary.tex
\section{Preliminary}
\label{sec:preliminary}


\paragraph{Program Reproduction Task.}
Software engineering agents are increasingly expected to develop complete software systems with limited human supervision~\citep{yang2026programbench}. Issue-resolution tasks such as SWE-bench, however, start from an existing codebase and ask the agent to make localized changes~\citep{jimenez2024swe}. Program reproduction from scratch removes this implementation scaffold and poses a more demanding task. Given only a software program and its documentation, the agent must produce source code that recovers the program's functionality while independently making the software design decisions needed to architect~\citep{yang2026programbench}. Training agents on this task requires stable environments that support behavioral exploration and automatically assess candidate implementations. Let $P_i$ denote a source-available codebase drawn from a repository. We define an environment-construction pipeline $\Phi$ as:
\begin{equation}
    e_i = \Phi(P_i) = (B_i^\star, D_i, \mathcal{T}_i,\mathcal{S}_i),
    \qquad
    \mathcal{E} = \{e_i\}_{i=1}^{N}.
    \label{eq:agentic_environment_scaling}
\end{equation}
Here, $\mathcal{E}$ denotes the collection of $N$ program-reproduction environments constructed by $\Phi$. Each environment $e_i$ comprises an execute-only reference program $B_i^\star$ built from $P_i$, its usage documentation $D_i$, a black-box behavioral test suite $\mathcal{T}_i$, and a network-isolated runtime sandbox $\mathcal{S}_i$ in which the task agent operates. The agent can use any compiled language  (e.g., C/C++, Golang, Rust, Java) and must provide source code and a compile script that produces a candidate executable program $\widehat{B}_i$. The test suite $\mathcal{T}_i$ compares the externally observable behavior of $\widehat{B}_i$ and $B_i^\star$ on selected inputs, and its outcomes form the basis for behavior-grounded feedback during training. Because $\mathcal{T}_i$ covers only finitely many inputs, passing it establishes agreement on the tested behaviors rather than complete semantic equivalence~\citep{yang2026programbench}. Scaling this construction across diverse repositories poses two challenges: (i) reliably building reference programs with a unified pipeline despite variation in implementation languages, dependencies, and build procedures; and (ii) synthesizing finite black-box test suites with sufficient behavioral coverage and distinguishing faithful reproductions from incorrect ones.


%% file: section/3_method.tex

\section{Program-Env}\label{sec:program-env}
\begin{figure}[!htbp] 
    \centering
    
    \includegraphics[width=.95\linewidth]{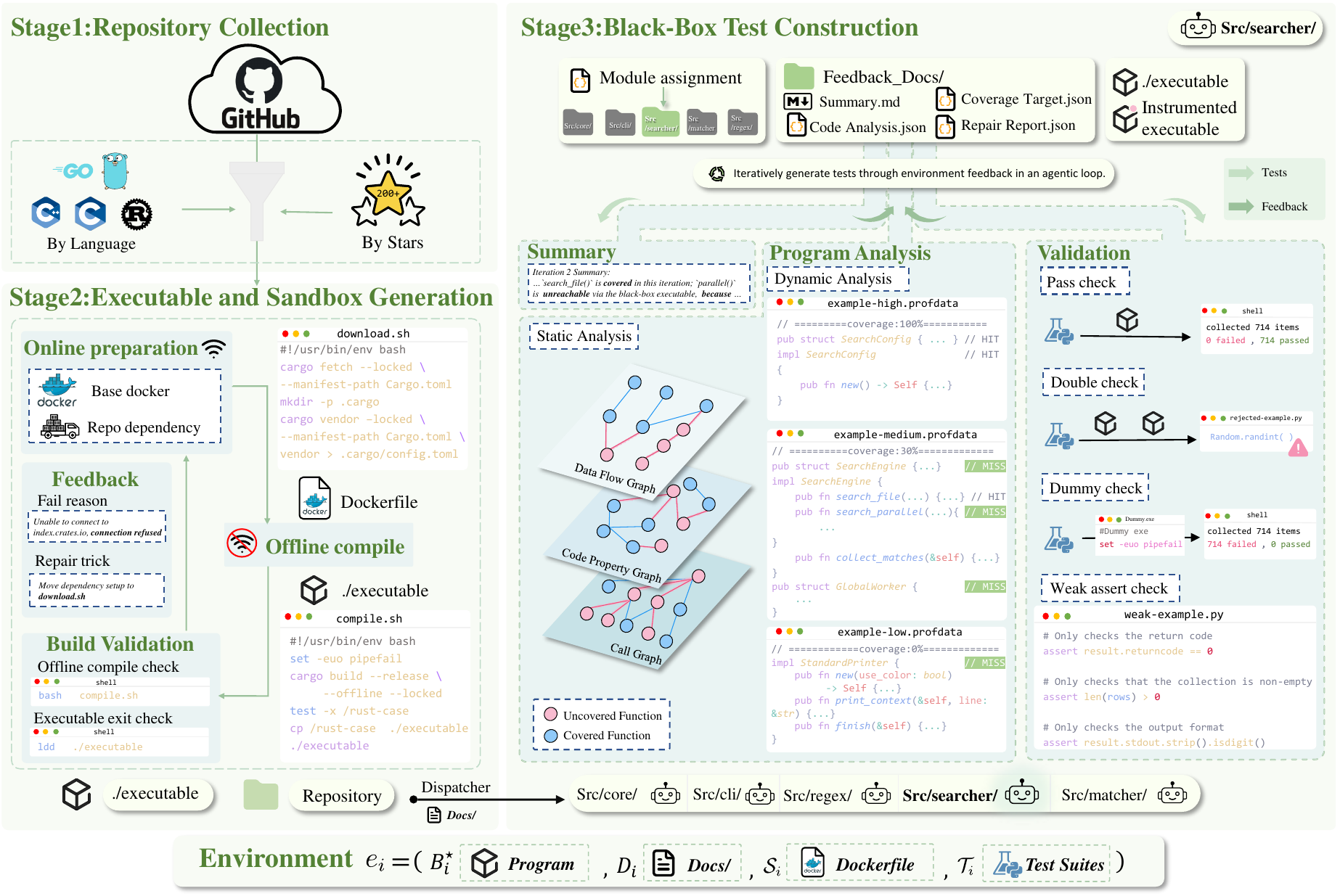} 
    \caption{Overview of the Program-Env. Stage~1 collects repositories from GitHub by language and stars. Stage~2 builds the gold executable $B_i^\star$ and runtime sandbox $\mathcal{S}_i$. Bridging the two stages, the dispatcher extracts documentation $D_i$ and splits the source into modules. In Stage~3, each module is assigned to a subagent that iteratively writes the test suite $\mathcal{T}_i$ under feedback.}
    \label{fig:program-env}
\end{figure}


Constructing program-reproduction environments at scale requires reliable executable rebuild from diverse repositories and comprehensive black-box behavioral tests that broadly capture program functionality. To overcome these challenges, we introduce \textbf{Program-Env}, which uses general harnesses such as \texttt{Claude Code} and \texttt{Codex} and adds feedback loops to guide the overall process of build repair and test refinement. As illustrated in Figure~\ref{fig:program-env}, Program-Env first collects candidate repositories $\{P_i\}_{i=1}^{M}$ (Section~\ref{sec:repo_collection}) and then constructs each retained environment $e_i$ by building the reference executable and runtime sandbox (Section~\ref{sec:build}) and generating black-box tests (Section~\ref{sec:tests}).

\subsection{Repository Collection}\label{sec:repo_collection}




To construct a diverse candidate pool of repositories for our program reproduction gym, Program-Env collects repositories through a stratified search process using the GitHub Search API. Specifically, we partition search queries by programming language and star-count buckets, while excluding archived and forked repositories. We further validate repository metadata, remove duplicate entries, and retain those with valid licensing metadata. Repositories already used by existing code-related benchmarks are also excluded. Each retained repository is anchored to a fixed commit and screened for explicit build-configuration evidence. This process yields $M{=}30{,}459$ candidate repositories $\{P_i\}_{i=1}^{M}$. More details are provided in Appendix~\ref{Appendix_repo_collection}.

\subsection{Executable and Sandbox Generation}\label{sec:build}



Program reproduction from scratch requires a prepared runtime sandbox with the necessary toolchains and dependencies so that the agent can focus on rebuilding program functionality. To accommodate differences across repositories, Program-Env pipeline $\Phi$ uses \texttt{Claude Code} as the underlying agent harness and adds an outer loop that validates each proposed build procedure and provides the diagnostics feedback for revision. For each repository $P_i$ with a standalone executable target, the agent generates \texttt{download.sh} for online dependency preparation and \texttt{compile.sh} for offline compilation. The overall loop proceeds to the next stage only if both scripts succeed and \texttt{./executable} passes a loadability check. Otherwise, failure diagnostics are returned to the agent for revision. The validated executable serves as $B_i^\star$, and the prepared image provides the basis for the runtime sandbox $\mathcal{S}_i$. Further details are provided in Appendix~\ref{Appendix:build}.

Finally, once the build validation succeeds, the dispatcher prepares the resources for black-box test construction. It uses codebase size to determine the number of parallel branches and assigns each branch a functional module. Alongside module assignment, it instruments the executable for coverage measurement and prepares the test runtime. It also extracts repository documentation such as \texttt{README.md} to form $D_i$. These preparation steps run in parallel as detailed in Appendix~\ref{Appendix:dispatcher}.

\subsection{Black-Box Tests Construction}\label{sec:tests}

Reliable verification is essential for effective supervision in agent training. For program reproduction tasks, a black-box test suite $\mathcal{T}_i$ provides this supervision by comparing $\hat{B}_i$ against $B_i^\star$ on shared inputs. However, tests generated in a single attempt may cover too few behaviors or use assertions that fail to detect incorrect implementations. Program-Env therefore uses structured feedback to guide the agent in iteratively expanding test coverage and strengthening assertions.

Following the dispatcher's module assignments, test construction proceeds in parallel across branches. Each branch runs the \texttt{Claude Code} harness inside an outer loop that validates each delivered suite and returns diagnostics for revision. After each round, the pipeline organizes the feedback into three components, stored as structured documents in \texttt{Feedback\_Docs/}.
(i) \textbf{Summary} records what was done and what remains unsolved and why, preserving context across rounds.
(ii) \textbf{Program analysis} locates uncovered functionality with dynamic analysis and traces it to public entry points with static analysis.
(iii) \textbf{Validation} screens tests through four checks: the pass check identifies tests that fail on $B_i^\star$; the double check detects inconsistent outcomes across repeated runs; the dummy check identifies tests that accept a dummy implementation; and the weak-assert check flags insufficient assertions. These checks help identify invalid, unstable, or insufficiently discriminative tests. In the next round, the agent reads these processed documents to repair flagged tests before extending coverage to unexplored targets. A subagent terminates once its module reaches the functionality coverage threshold or the round limit. Once all branches terminate, their suites are merged and validated again, and $\mathcal{T}_i$ retains only the tests that pass all checks. Further details are provided in Appendix~\ref{Appendix:tests}.


\section{Program-Agent}\label{sec:program-agent}
\begin{figure}[!htbp] 
    \centering
    \includegraphics[width=.68\linewidth]{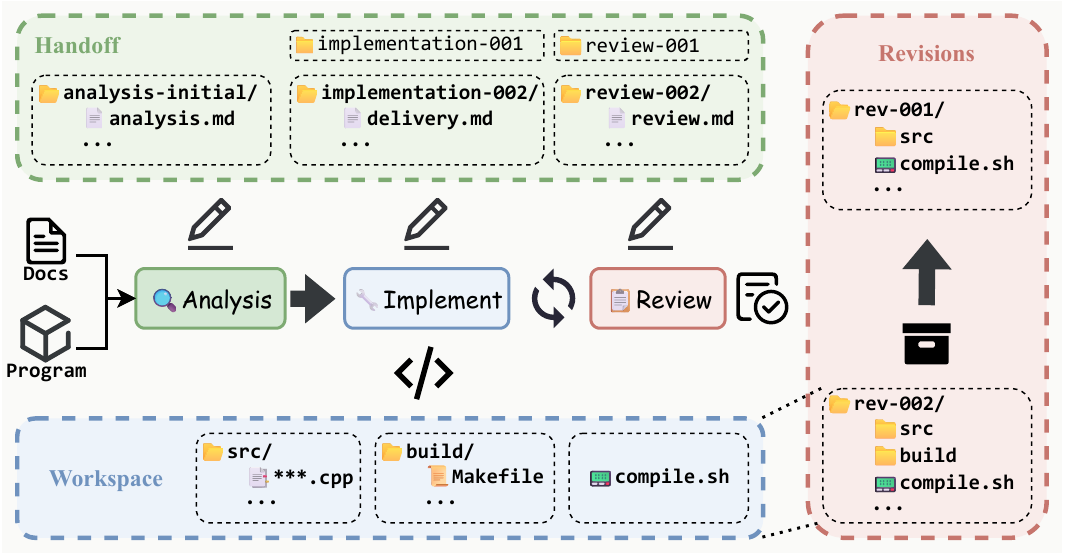} 
    \caption{Overview of Program-Agent. The \textbf{Analysis} stage explores the usage documentation and reference program to identify behavioral requirements. The \textbf{Implement} and \textbf{Review} stages then alternate to refine the candidate implementation for a fixed loop budget. \protect\textcolor[HTML]{7DAA72}{Handoff} carries behavioral evidence and delivery reports across isolated sessions, \protect\textcolor[HTML]{6F92BE}{Workspace} maintains the evolving implementation, and \protect\textcolor[HTML]{C6756D}{Revisions} preserves independently buildable source code for validation and recovery.}
    \label{fig:program-agent}
\end{figure}

In long-horizon program reproduction, single-session workflows may lose early behavioral evidence due to context compression and consequently produce a program that fails to compile~\citep{claude_ce,hao2026self,feng2026longcli}. Inspired by the iterative workflow of software engineers, we propose \textbf{Program-Agent} built on \texttt{Claude Code} to support systematic development in long-horizon program reproduction. As illustrated in Figure~\ref{fig:program-agent}, it combines an iterative role-isolated workflow (Section~\ref{subsec:role-isolated-workflow}) with a revision management mechanism (Section~\ref{subsec:validated-revision-management}) for compilation validation and failure recovery.


\subsection{Role-Isolated Iterative Workflow}
\label{subsec:role-isolated-workflow}
Program-Agent adopts a role-isolated and session-separated iterative paradigm. It decouples program reproduction into two logical roles (analyst and implementer) and three stages: \textbf{Analysis}, \textbf{Implement}, and \textbf{Review}. Each stage executes in an independent session to eliminate cross-stage memory contamination. Handoff artifacts carry behavioral evidence and stage reports across sessions without transferring the full conversation history.

\paragraph{Analysis Stage.}
The analyst reads the documentation and explores the reference program, without editing Workspace, guaranteeing unbiased constraint collection. All extracted behavioral evidence and implementation guidelines are stored in \texttt{analysis-initial/analysis.md} and frozen as read-only reference for all downstream stages.

\paragraph{Implement Stage.}
The implementer develops candidate programs in an isolated session based on frozen Handoff artifacts without accessing the reference program, to avoid context noise. The first iteration delivers a buildable baseline to avoid fundamental compilation failures, while subsequent rounds incrementally resolve verified review issues. All implementation progress and code changes are documented in \texttt{implementation-k/delivery.md} for explicit traceability.

\paragraph{Review Stage.}
The analyst compares the official reference program with the newly built, immutable candidate program, and converts all verified behavioral discrepancies into actionable revision requirements recorded in \texttt{review-k/review.md}. Strict read-only access is enforced on both the candidate program and the Workspace to prevent self-justifying modifications.

\paragraph{Iteration budget.}
Given a fixed budget of $K$ implementation rounds, Program-Agent executes one standardized Analysis stage followed by alternating Implement and Review iterations. This explicit iteration budget resolves two common issues in single-session workflows: premature termination with incomplete implementations, and unbounded low-value exploration that destabilizes long-horizon reproduction. 

\subsection{Revision Management}
\label{subsec:validated-revision-management}
Across multiple Implement rounds, the shared Workspace can accumulate error edits until \texttt{compile.sh} no longer produces an executable. Program-Agent therefore treats independent compilation as the gate for each implementation round. It exports a clean snapshot of the committed source and compiles that snapshot outside the development workspace. A snapshot that produces an executable is archived in Revisions and forwarded to the next Review. Moreover, a stage that times out or misses required artifacts or fails this compilation is retried in a fresh session within a bounded budget. If the retry still fails, the implementation rolls back to the last validated revision. A successful revision becomes the recovery checkpoint for later rounds. This keeps progress stable across multi-stage reproduction. More details are provided in Appendix~\ref{app:program-agent}.





%% file: section/4_experiment.tex
\section{Experiments and Analysis}\label{sec:experiments}


This section presents an evaluation of ProgramGym. Our experiments are designed to answer the following four research questions. \textbf{RQ1 (Effectiveness of ProgramGym)} Does training on Program-Env and solving with Program-Agent improve program reproduction task? \textbf{RQ2 (Training Environments Analysis)} What scale and construction dynamics characterize the environments produced by Program-Env? \textbf{RQ3 (Test-Time Scaling)} Does additional Program-Agent iteration budget improve Tests Passed? \textbf{RQ4 (Agent Behavior Analysis)} How does ProgramGym influence agent behavior on program reproduction tasks?

\subsection{Experimental Settings}
\paragraph{Environment Statistic.}
Program-Env constructs 1,648 training environments spanning C, C++, Go, and Rust, using \texttt{DeepSeek-V4-Flash-0731}~\citep{deepseekai2026deepseekv4} as the underlying model. Figure~\ref{fig:image_dataset_overview_repo} shows that the repositories fall into eight functional categories and that developer tools and CLI \& system utilities are the two largest. Test construction uses 5,918 subagents across these repositories. Table~\ref{tab:planning-input-categories} reports six black-box interaction categories. We then collect trajectories for subsequent training by rejection sampling \texttt{DeepSeek-V4-Flash-0731}~\citep{deepseekai2026deepseekv4} with \texttt{Claude Code} and filtering on test-case pass rate.
\input{table/overview_static}

\paragraph{Agent Scaffolds.}
To support interactive program reproduction from scratch, we use \texttt{Claude Code} rather than \texttt{mini-SWE-agent} as the single-session baseline. The task is long-horizon and often exceeds a $256$K maximum context window. \texttt{Claude Code} compresses context so the agent can continue, following the ProgramBench protocol of GLM-5.2~\citep{glm5team2026glm5vibecodingagentic}. We compare it with Program-Agent using the same container resource limits. Program-Agent runs for up to $K{=}5$ iterations. Each task runs in an isolated container with 8 CPU cores, 16\,GB of RAM, and 30\,GB of storage, under an 8-hour wall-clock limit.

\paragraph{Models and Training Details.}
We fine-tuned \texttt{Qwen3.8-27B}~\citep{qwen38} on the collected trajectories with the ms-swift framework~\citep{zhao2024swiftascalablelightweightinfrastructure} and obtain \texttt{Qwen3.8-27B-SFT}. We evaluate both checkpoints on ProgramBench and use \texttt{DeepSeek-V4-Flash-0731} (\texttt{DS-V4-Flash-0731})~\citep{deepseekai2026deepseekv4} to further assess our proposed Program-Agent across model backbones. More details are provided in Appendix~\ref{app:training-details}.

\paragraph{Evaluation Metrics.}
We follow the evaluation metrics of ProgramBench~\citep{yang2026programbench}. \textbf{\% Tests Passed} is the macro-average of per-task test pass rates. \textbf{\% Almost Resolved} is the percentage of tasks with at least $95\%$ of tests passing. \textbf{\% Resolved} is the percentage of tasks that pass all tests without a cheating flag.

\subsection{Effectiveness of ProgramGym (\textbf{RQ1})}

To verify the effectiveness of our proposed \textbf{ProgramGym}, we evaluate \texttt{DS-V4-Flash-0731}, \texttt{Qwen3.8-27B}, and \texttt{Qwen3.8-27B-SFT} on ProgramBench and compare them with frontier models that use \texttt{mini-SWE-agent} (Table~\ref{tab:main-results}).
\begin{table}[t]
    \centering
    \scriptsize
    
    \caption{Results on ProgramBench across models and agent scaffolds. Program-Agent uses $K{=}5$ implementation rounds. $\Delta$ is relative to the base model with \texttt{Claude Code}.}
    \label{tab:main-results}
    \renewcommand{\arraystretch}{0.9}
    \begin{tabular}{lcc ccc cc}
        \toprule
        \multicolumn{1}{c}{\multirow{2}{*}{Models}}
        & \multicolumn{1}{c}{\multirow{2}{*}{Agent Scaffolds}}
        & \multicolumn{2}{c}{\textbf{Tests Passed}}
        & \multicolumn{2}{c}{\textbf{Almost Resolved}}
        & \multicolumn{2}{c}{\textbf{Resolved}} \\
        & & (\%, $\uparrow$) & $\Delta$
        & (\%, $\uparrow$) & $\Delta$
        & (\%, $\uparrow$) & $\Delta$
        \\
        \midrule
        \rowcolor{lightgray}
        \multicolumn{8}{c}{\textbf{Flagship Models}~\citep{yang2026programbench}} \\
        \midrule
        GPT 5 mini              & mini-SWE-agent & 16.00 & -- & 0.00 & -- & 0.00 & -- \\
        GPT 5.4 mini            & mini-SWE-agent & 16.40 & -- & 0.00 & -- & 0.00 & -- \\
        Claude Haiku 4.5        & mini-SWE-agent & 30.00 & -- & 0.00 & -- & 0.00 & -- \\
        Gemini 3 Flash          & mini-SWE-agent & 31.80 & -- & 0.00 & -- & 0.00 & -- \\
        Gemini 3.1 Pro          & mini-SWE-agent & 36.40 & -- & 0.00 & -- & 0.00 & -- \\
        GPT 5.4                 & mini-SWE-agent & 37.70 & -- & 0.00 & -- & 0.00 & -- \\
        Claude Sonnet 4.6       & mini-SWE-agent & 47.50 & -- & 1.00 & -- & 0.00 & -- \\
        Claude Opus 4.7         & mini-SWE-agent & 50.90 & -- & 3.00 & -- & 0.00 & -- \\
        Claude Opus 4.6         & mini-SWE-agent & 52.10 & -- & 2.50 & -- & 0.00 & -- \\
        Gemini 3.5 Flash        & mini-SWE-agent & 53.60 & -- & 3.00 & -- & 0.00 & -- \\
        Claude Opus 4.7 (xhigh) & mini-SWE-agent & 55.10 & -- & 4.50 & -- & 0.00 & -- \\
        \midrule
        \rowcolor{lightgray}
        \multicolumn{8}{c}{\textbf{DeepSeek-V4-Flash-0731}~\citep{deepseekai2026deepseekv4}} \\
        \midrule
        DS-V4-Flash-0731   & Claude Code   & 57.68 & -- & 4.50 & -- & 0.00 & --\\
        DS-V4-Flash-0731   & Program-Agent & 58.97 & 1.29 & 5.0 & 0.5 & 0.00 & 0.00 \\
        \midrule
        \rowcolor{lightgray}
        \multicolumn{8}{c}{\textbf{Qwen3.8-27B}~\citep{qwen38}} \\
        \midrule
        Qwen3.8-27B   & Claude Code   & 34.87 & -- & 3.50 & -- & 0.50 & -- \\
        Qwen3.8-27B   & Program-Agent & 55.71 & 20.84 & 5.50 & 2.00 & 0.50 & 0.00  \\
        Qwen3.8-27B-SFT & Claude Code   & 42.84 & 7.97 & 1.50 & -2.00 & 0.00 & -0.50  \\
        Qwen3.8-27B-SFT & Program-Agent & 56.25 & 21.38 & 4.50 & 1.0 & 0.50 & 0.00  \\
        \bottomrule
    \end{tabular}
\end{table}

\textbf{Program-Env scales verifiable environments for program reproduction from scratch.} Holding \texttt{Claude Code} fixed keeps the agent scaffold unchanged and varies only whether the model was trained on Program-Env trajectories. On ProgramBench, \texttt{Qwen3.8-27B-SFT} reaches 42.84\% Tests Passed against 34.87\% for \texttt{Qwen3.8-27B} ($\Delta{=}+7.97$) and narrows the gap to \texttt{DS-V4-Flash-0731} under the same agent scaffold (57.68\%). The \texttt{Qwen3.8-27B-SFT} also exceeds the flagship \texttt{GPT 5.4} result (37.70\%) under \texttt{mini-SWE-agent}. These results indicate that Program-Env's runtime sandboxes and black-box tests provide reliable behavioral supervision for training on program reproduction tasks.

\textbf{Program-Agent expands a model's capability boundary for long-horizon program reproduction tasks.} Holding the model fixed keeps the weights unchanged and varies only the agent scaffold. Program-Agent raises \texttt{Qwen3.8-27B} from 34.87\% to 55.71\% Tests Passed and \texttt{Qwen3.8-27B-SFT} from 42.84\% to 56.25\% relative to \texttt{Claude Code}. Each run exceeds \texttt{Claude Opus 4.7} (50.90\%) and \texttt{Gemini 3.5 Flash} (53.60\%) under \texttt{mini-SWE-agent}. These results indicate that the software-engineer workflow recovers more program behavior from the same model than a single-session scaffold.

\textbf{Fully reproducing a program remains hard.} Tests Passed can exceed 50\% while Almost Resolved and Resolved stay near zero. Flagship models under \texttt{mini-SWE-agent} remain at 0.00\% Resolved, and after SFT Almost Resolved falls from 3.50\% to 1.50\% under \texttt{Claude Code}. These results indicate that whole-program reproduction from scratch is still unsolved.

\subsection{Training Environments Analysis (\textbf{RQ2})}

\begin{figure}[h] 
    \centering
    
    \includegraphics[width=1.\linewidth]{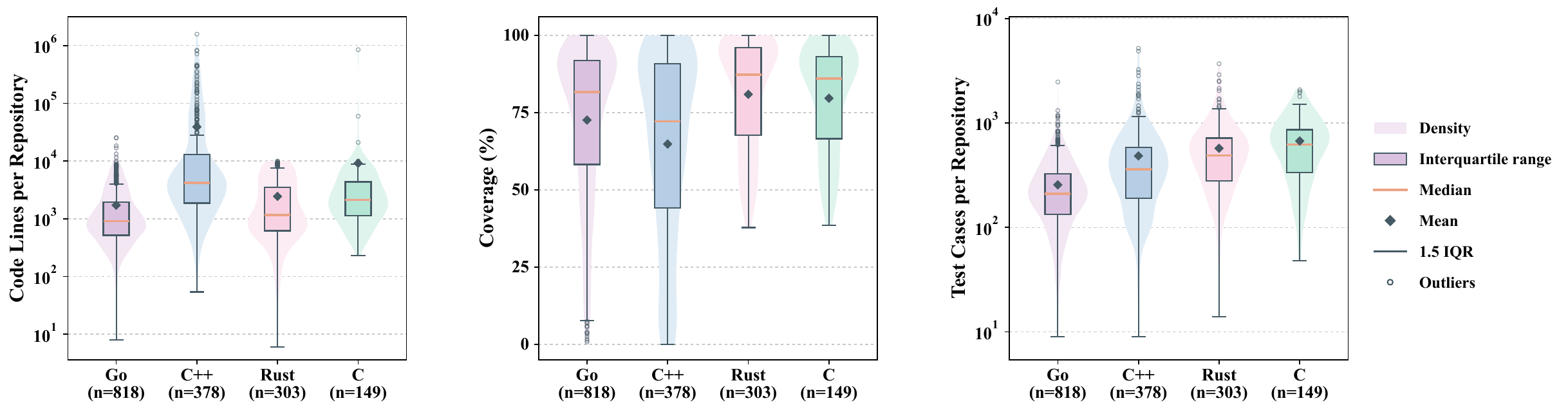} 
    \caption{Distributions of code lines, coverage, and test cases in Program-Env.} 
    
    
    \label{fig:image_dataset_boxplot}
\end{figure}

Program-Env comprises 1,648 tasks built from open-source repositories spanning eight functional categories (Figure~\ref{fig:image_dataset_overview_repo}), at a total cost of 136.77B tokens (Figure~\ref{fig:dataset_overview_token}a). The environments cover four languages, C, C++, Go, and Rust, and range from small utilities to large, complex projects (Figure~\ref{fig:image_dataset_boxplot}).

To make verification reliable, the feedback loop iteratively expands coverage and refines the test suite, yielding a mean coverage of 72.99\% (median 81.63\%) and 665,512 validated test cases in total (Figure~\ref{fig:dataset_overview_token}b,c), so the suites capture most of each repository's functionality. Figure~\ref{fig:image_dataset_boxplot} illustrates both points at the repository level. (i)\textbf{ Coverage} clusters tightly near 80\% with a long lower tail, where the low-coverage repositories are those whose code is only partially reachable through the black-box interface, which does not affect the validity of $\mathcal{T}_i$. (ii) \textbf{Test suite} sizes are concentrated around each language's mean, with an overall mean of 404 cases per repository. Program-Env thus provides a reliable, diverse, and scalable training environment for program reproduction from scratch. Further descriptive statistics are given in Appendix~\ref{appendix:Statistics}.

\subsection{Test-Time Scaling (\textbf{RQ3})}

\textbf{Program-Agent can scale performance with extra iteration budget.} We evaluate \texttt{Qwen3.8-27B-SFT} with Program-Agent under iteration budgets (Figure~\ref{fig:plot_agent}). Tests Passed rises from 43.49\% at $K{=}1$ to 54.13\% at $K{=}3$ and 56.25\% at $K{=}5$. Five iterations are comparable to \texttt{DS-V4-Flash-0731} (57.68\%). Most of the gain arrives by $K{=}3$. These results indicate that Program-Agent keeps recovering program behavior as the iteration budget grows. Extra analysis about costs is given in Appendix~\ref{app:token-cost-analysis}.

\input{table/overview_agent}
\subsection{Agent Behavior Analysis (\textbf{RQ4})}

\textbf{ProgramGym changes agent behavior mainly through Program-Agent.} We compare action types of \texttt{Qwen3.8-27B} and \texttt{Qwen3.8-27B-SFT} under \texttt{Claude Code} and Program-Agent (Figure~\ref{fig:tool_use}). Both models use mainly execute and probe under \texttt{Claude Code}. Program-Agent increases read and git, most after Program-Env training. These results indicate that Program-Agent reshapes the agent toward exploration and revision. Further analysis is given in Appendix~\ref{app:extra-analysis-of-agent-behavior}.

%% file: table/overview_static.tex
\begin{figure*}[t]
\centering

\begin{minipage}[t]{0.42\textwidth}
    \vspace{0pt}
    \centering
    \includegraphics[width=0.85\linewidth]{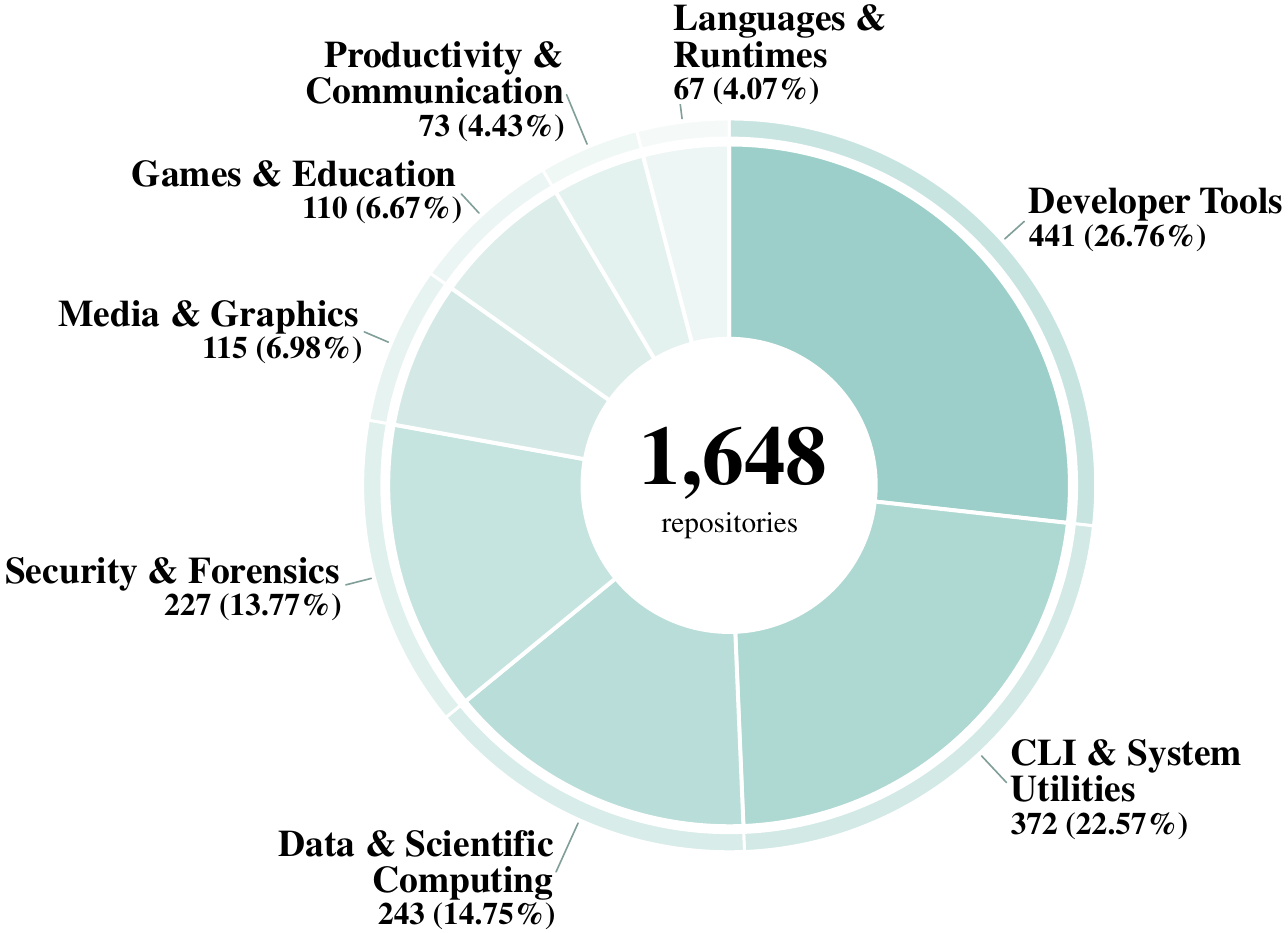}

    \captionof{figure}{
        Distribution of the 1,648 repositories
        across eight functional categories.
    }
    \label{fig:image_dataset_overview_repo}
\end{minipage}
\hfill
\begin{minipage}[t]{0.52\textwidth}
    \centering
    \small
    \renewcommand{\arraystretch}{1.15}
    \captionof{table}{
        Distribution of the black-box interaction categories over 5,918 subagents from 1,648 repositories.
    }
    \begin{tabular*}{\linewidth}
        {@{\extracolsep{\fill}}lrr@{}}
        \toprule
        \textbf{Tests Category}
        & \textbf{\# Subagents}
        & \textbf{\% Percentage} \\
        \midrule
        Arguments     & 4,901 & 82.82\% \\
        Configuration & 2,388 & 40.35\% \\
        Help          & 4,070 & 68.77\% \\
        I/O           & 5,873 & 99.24\% \\
        Subcommands   & 1,305 & 22.05\% \\
        TUI           &   716 & 12.10\% \\
        \bottomrule
    \end{tabular*}

    \label{tab:planning-input-categories}
\end{minipage}

\end{figure*}


%% file: table/overview_agent.tex
\begin{figure*}[h]
\centering

\begin{minipage}[t]{0.32\textwidth}
    \vspace{0pt}
    \centering
    \includegraphics[width=0.90\linewidth]{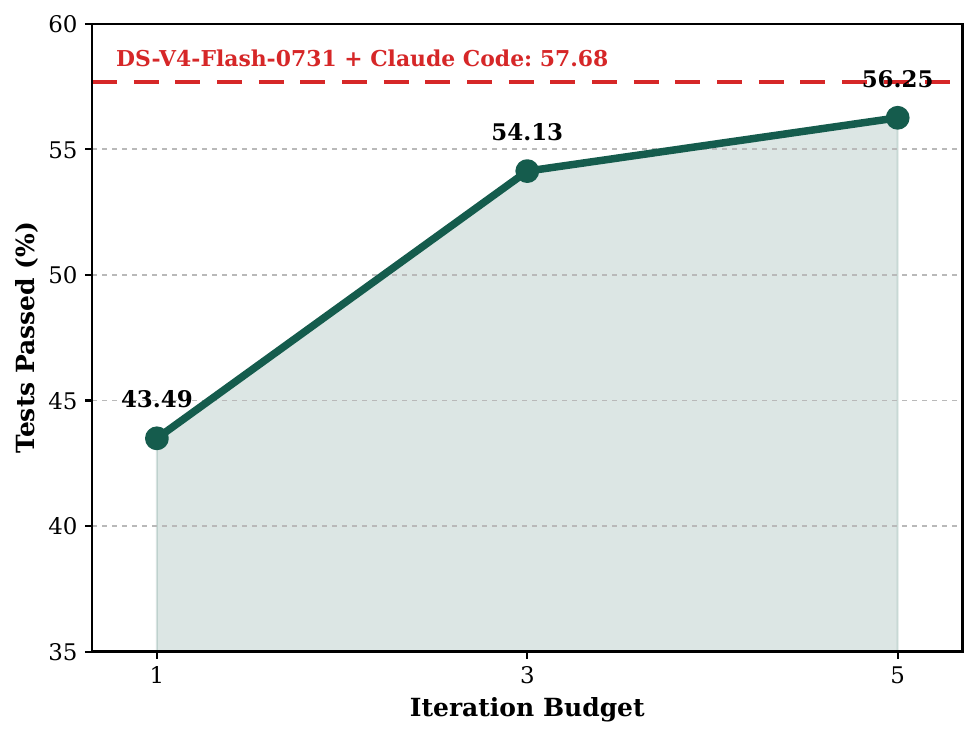}

    \captionof{figure}{
        Tests Passed of \texttt{Qwen3.8-27B-SFT} with Program-Agent across different iteration budgets on ProgramBench.
    }
    \label{fig:plot_agent}
\end{minipage}
\hfill
\begin{minipage}[t]{0.65\textwidth}
    \vspace{0pt}
    \centering
    \includegraphics[width=1.0\linewidth]{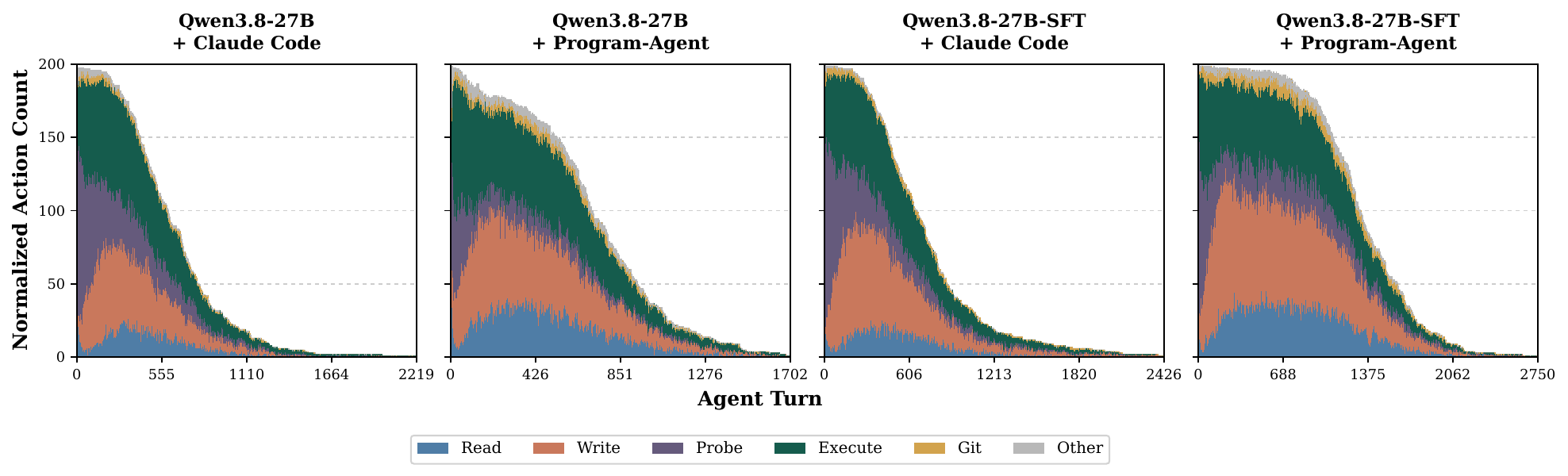}

    \captionof{figure}{
   Distribution of action types over agent turns. Each task contributes a total weight of one per turn, divided evenly among its tool calls and aggregated by action type. Bar heights decay with turn index because trajectories end at different lengths.
    }
    \label{fig:tool_use}
\end{minipage}

\end{figure*}


%% file: section/5_discussion.tex

%% file: section/6_conclusion.tex
\section{Conclusion}
In this paper, we propose \textbf{ProgramGym} for training agents on long-horizon program reproduction from scratch. Program-Env is a feedback-driven pipeline that builds the reference executable and runtime sandbox and constructs black-box tests from repositories and yields $1{,}648$ training environments in C, C++, Go, and Rust. Program-Agent combines a role-isolated workflow of analysis, implementation, and review with revision management for compilation checks and failure recovery. Fine-tuning on trajectories with these constructed environments raises \texttt{Qwen3.8-27B} from $34.87\%$ to $42.84\%$ on ProgramBench. Program-Agent further raises the base and SFT models to $55.71\%$ and $56.25\%$, both exceeding some flagship models. Extra Program-Agent iterations continue to recover program behavior and Program-Agent rather than SFT alone reshapes the agent toward exploration and revision. We will release ProgramGym as a scalable gym for the community to develop coding agents that discover requirements and design systems from scratch.

%% file: extra.tex
\subsection*{AI use statement}

In this work, we used generative AI tools to implement methods in our experiments. Program-Env and Program-Agent use agent harnesses such as \texttt{Claude Code} to construct environments, generate tests, and collect trajectories, as described in Sections~\ref{sec:build} and~\ref{sec:tests}. Additionally, we used generative AI tools to edit the paper to improve readability. We reviewed all AI-assisted edits. We take responsibility for the final content of this work, including text, claims, and artifacts produced with the aid of generative AI.

\subsection*{Reproducibility statement}

We document the materials needed to reconstruct ProgramGym. Repository collection, commit locking, and filtering strategies are specified in Section~\ref{sec:repo_collection} and Appendix~\ref{Appendix_repo_collection}. Executable and sandbox construction, including online preparation, offline compilation, and build-stage validation, are described in Section~\ref{sec:build} and Appendix~\ref{Appendix:build}. Dispatcher modules, coverage instrumentation, and test-generation sandboxes appear in Appendix~\ref{Appendix:dispatcher}. Black-box test construction, the four validation checks, and the feedback loop are given in Section~\ref{sec:tests} and Appendix~\ref{Appendix:tests}. Program-Agent's role-isolated workflow, isolation rules, and recovery checks are described in Section~\ref{subsec:role-isolated-workflow} and Appendix~\ref{app:program-agent}. Training hyperparameters are listed in Appendix~\ref{app:training-details}. Evaluation uses ProgramBench with the metrics and container limits in the experimental settings. We will release ProgramGym, including environments, construction code, and training artifacts, to support reproduction.

%% file: section/8_Appendix.tex
\begin{center}
{\LARGE\bfseries Appendix}
\end{center}

\vspace{1.2em}
\noindent\textsc{Contents}
\vspace{0.8em}

\setcounter{tocdepth}{2}
\makeatletter
\renewcommand*{\l@section}[2]{%
  \@dottedtocline{1}{0em}{2.3em}{\bfseries #1}{\bfseries #2}}
\renewcommand*{\l@subsection}{\@dottedtocline{2}{1.5em}{3.0em}}
\renewcommand*{\l@subsubsection}{\@dottedtocline{3}{3.8em}{4.2em}}
\@starttoc{toc}
\makeatother

\clearpage
\section{FREQUENTLY ASKED QUESTIONS (FAQs)}\label{FAQ}

\paragraph{Why are there far fewer tasks than collected repositories?}
Two factors account for the gap between the 30,459 collected repositories and the 1,648 tasks. (i) \textbf{Filtering.} Many candidates cannot serve as black-box tasks because they fail to build on Linux, require a display or user input devices, or exceed the 3,600-second limit for compilation or instrumentation, which we cannot leave unbounded. Such repositories are excluded before the test construction stage. (ii) \textbf{Cost.} Constructing an environment is expensive, about 82.99M tokens on average (Appendix~\ref{appendix:Statistics}), so processing the full candidate pool was beyond our budget. The pipeline itself, however, is fully automated, so that the community can construct further high-quality environments from the remaining repositories.

\paragraph{Why is the difficulty distribution skewed toward easy tasks?}
Under the complexity score $\xi$ of Section~\ref{appendix:decomposition}, the 1,648 tasks split into 1,492 easy, 151 medium, and 5 hard. This skew is intentional. (i) $\xi$ measures repository size rather than the difficulty of program reproduction from scratch: a large codebase may expose few black-box entry points and little observable behavior. (ii) Tasks far beyond current model capability yield almost no successful trajectories, so collecting more of them adds little effective training signal. (iii) Despite the skew, models trained on Program-Env show clear gains in program reproduction ability (Section~\ref{sec:experiments}). As model capability grows, the same pipeline can be scaled to medium and hard repositories to push the frontier of coding models further.

\paragraph{Why not report a mean coverage curve over iterations?}
We could not report a mean coverage curve over iterations for two reasons. 
(i) \textbf{Subagents terminate at different rounds}, whether by reaching the coverage threshold, the round limit, or early stopping (Appendix~\ref{appendix:loop-termination}), so averaging over surviving modules would be biased toward the hardest ones and understate the gain from feedback. 
(ii) \textbf{Coverage is only aggregated to the repository level after all branches terminate}, so no repository-level curve over iterations exists. We therefore present case studies instead. The two cases in Figure~\ref{fig:test_retry}
were chosen to represent the two effects of feedback, repairing weak tests and expanding functionality coverage. 

\paragraph{Why can the Implement stage of Program-Agent not access the reference program?}
This restriction follows the separation of responsibilities in software engineering: Analysis extracts behavioral requirements, Implement produces a buildable source revision from Handoff artifacts, and Review independently identifies remaining discrepancies. If the Implement stage could access the reference program, the agent might continue probing and matching marginal edge cases after completing its primary objectives, delaying delivery and degrading the staged workflow into open-ended trial and error. Program-Agent instead delegates further behavioral exploration to Review, which converts verified discrepancies into explicit requirements for the next round. This design preserves clear stage boundaries while ensuring steady and auditable progress.

\paragraph{Why do we not compare with MindForge?}
MindForge~\citep{chen2026mindforgeteachingsmalllanguage} has not released its environments or trajectories. Without these artifacts, we cannot evaluate under a matched protocol, so we do not report a head-to-head comparison.

\section{Program-Env Details}\label{app:program-env}
As illustrated in Figure~\ref{fig:program-env}, Program-Env is constructed through a multi-stage pipeline, and this section provides the implementation details of each stage. We first describe how repositories are collected and filtered in Section~\ref{Appendix_repo_collection}. Section~\ref{Appendix:build} then explains how each repository is built into a validated executable $B_i^\star$ and runtime sandbox $\mathcal{S}_i$. Building on these validated builds, Section~\ref{Appendix:dispatcher} presents the dispatcher, which transforms each built repository into the resources required for test generation through task decomposition, coverage instrumentation, test-generation sandbox construction, and documentation extraction, producing the test-generation sandbox $\mathcal{S}_i^{\mathrm{test}}$ and documentation $D_i$. Finally, Section~\ref{Appendix:tests} details how parallel subagents iteratively construct, validate, and refine the final test suite $\mathcal{T}_i$ under execution, coverage, and program-analysis feedback.

\subsection{Repository Collection Details}\label{Appendix_repo_collection}
In this section, we describe our four-stage pipeline for systematically collecting and filtering open-source repositories from GitHub. The pipeline consists of Comprehensive Search, Metadata Validation, Commit Locking, and Build Config Detection.

\paragraph{Comprehensive Search.}
For the corpus analyzed in this work, we search for repositories with C, C++, Go, or Rust as their primary language. Repository retrieval is conducted via the GitHub Search API. To circumvent its 1000-result limitation per query, we partition the search space independently for each language using six mutually exclusive star-count buckets: $[200,499]$, $[500,999]$, $[1{,}000,1{,}999]$, $[2{,}000,9{,}999]$, $[10{,}000,49{,}999]$, and $[50{,}000,\infty)$. If a query exceeds this limit, its star-count bucket is recursively bisected until the sub-query yields fewer than 1,000 results, the bucket can no longer be divided, or the maximum recursion depth of 12 is reached. For the open-ended highest bucket, the maximum star count observed in the returned results is used as a temporary upper bound for computing the bisection midpoint, while the resulting upper sub-query remains open-ended. This search strategy stratifies repositories by programming language and star count, improving repository coverage within the constraints of the GitHub Search API. Additionally, we include \texttt{fork:false} and \texttt{archived:false} in every query to reduce duplicate and inactive repositories in the candidate pool.

\paragraph{Metadata Validation.}
We locally re-validate the retrieved repository metadata to ensure consistency with the search criteria. Specifically, we check each repository's primary language and star count, remove duplicate repositories based on \texttt{full\_name}, and require a non-empty SPDX license identifier as a basic metadata completeness criterion. We further exclude repositories already used by existing code-related benchmarks, thereby preventing repository-level overlap with prior evaluation sets.

\paragraph{Commit Locking.}
Because GitHub repositories are continuously updated, we fix a specific code version for each qualified repository—specifically, the latest commit (i.e., the branch tip) of the default branch at the moment of data retrieval. For each candidate repository, we extract the \texttt{default\_branch} field from the repository metadata collected in the first stage, query the head commit of that branch and record its SHA. Ultimately, exactly one commit is retained for each repository.

\paragraph{Build Config Detection.}
After commit locking, we perform a lightweight, static detection of the selected commit to identify repositories with explicit build-configuration evidence. For each repository, we resolve the root tree of the selected commit and retrieve its recursive file tree through the GitHub REST API. We then search both the repository root and its subdirectories for common build-configuration markers relevant to target languages, including \texttt{Cargo.toml} for Rust, \texttt{go.mod} for Go, and native build-system files such as \texttt{CMakeLists.txt}, \texttt{Makefile}, and \texttt{meson.build} for C and C++. We additionally treat npm-related configuration, such as \texttt{package.json}, as auxiliary evidence of build infrastructure. A repository passes this stage if at least one recognized marker is found anywhere in its retrieved file tree; otherwise, it is excluded. This procedure serves only as a scalable proxy for the presence of build infrastructure and does not establish that a repository can be successfully compiled.

The final repository collection contains 30,459 repositories across C, C++, Go, and Rust. Detailed statistics for each stage of the collection pipeline are reported in Table~\ref{tab:repository_filtering}.

\begin{table}[t]
\centering
\caption{Repository statistics across the four-stage collection pipeline.}
\label{tab:repository_filtering}
\resizebox{\linewidth}{!}{%
\begin{tabular}{lrrrrrrr}
\toprule
\textbf{Stage} &
\textbf{Repositories} &
\textbf{Excluded} &
\textbf{Cumulative Retention} &
\textbf{C} &
\textbf{C++} &
\textbf{Go} &
\textbf{Rust} \\
\midrule
Comprehensive Search
& 41,833 & -- & 100.00\%
& 9,736 & 13,147 & 11,730 & 7,220 \\

Metadata Validation
& 35,710 & 6,123 & 85.36\%
& 7,604 & 10,720 & 10,777 & 6,609 \\

Commit Locking
& 35,710 & 0 & 85.36\%
& 7,604 & 10,720 & 10,777 & 6,609 \\

Build Config Detection
& 30,459 & 5,251 & 72.81\%
& 5,695 & 8,229 & 9,944 & 6,591 \\
\bottomrule
\end{tabular}%
}
\end{table}

\subsection{Build Procedure Details}\label{Appendix:build}

\paragraph{Online preparation and offline compile.}
Building a project requires a compiler, standard library, and build tools that match the project itself, and installing the latest version of each does not guarantee a successful build. As Table~\ref{tab:toolchain-mismatch} shows, the reasons a language breaks under its newest toolchain differ case by case. The build environment must therefore be prepared according to each project's actual version requirements rather than fixed per language.

To accommodate these differences, Program-Env equips the build agent with a skill for each supported language that encodes how to resolve toolchain versions, obtain dependencies, and drive the build. Guided by the skill, the agent makes environment preparation explicit in a script \texttt{download.sh}, which fetches and configures the required compiler, build tools, and third-party dependencies and exports the environment variables used by the subsequent build. A second script, \texttt{compile.sh}, then builds the gold executable program $B_i^\star$ within the prepared environment. The separation reflects the solving side, in which the source is reconstructed from interaction with the executable under no network access. Compiling offline verifies that the prepared environment is self-contained, sparing the solving model from generating missing dependencies or compilers.
\input{table/language_error}

\subsubsection{Build-stage Validation}\label{Appendix:build-validation}
\paragraph{Linter check.}
Validation combines a static linter with two-stage execution. The linter, shown in Table~\ref{tab:build-script-linter}, first ensures that the generated scripts do not mask failures, since a hidden error leaves nothing for the build agent to diagnose. It rejects \texttt{|| true}, redirections that discard standard error, and placeholder content, and requires acceptable failures to be handled by explicit conditional branches. The checks are intentionally strict, as a false positive costs a single repair round while an undetected failure propagates to every downstream stage.

\paragraph{Execution check.}
Execution then proceeds online and offline. The system first runs \texttt{download.sh} in a container with network access, providing the script alone and not providing the project source, so that the script must generate its own dependency manifest for prefetching. Upon success, the system records the exported environment variables and a snapshot of installed packages, and freezes the container state as the runtime sandbox $\mathcal{S}_i$. The system then places the project source in a separate working directory, launches a network-disabled container from $\mathcal{S}_i$, restores the recorded environment variables, and executes \texttt{compile.sh}. Since the build cannot reach the network, it succeeds only if the toolchain and local resources prepared in the first stage are sufficient. Validation checks the exit status of both scripts, the presence of an executable at the agreed location, and the format and linkage of that executable. As shown in Figure~\ref{fig:program-env}, the linkage check runs \texttt{ldd} on the $B_i^*$ to confirm that it was built successfully and no dynamic library is missing.

\clearpage
\input{table/compile_linter}

\subsubsection{Build-stage feedback}
\label{Appendix_feedback_build}

Our validation is deliberately strict to guarantee the stability of downstream stages, and current agents do not reliably satisfy every constraint in a single attempt (Figure~\ref{fig:build-feedback-heatmap}). The pipeline therefore carries each failure forward as structured feedback for the next attempt, comprising the previous \texttt{download.sh} and \texttt{compile.sh}, the diagnosis, and a suggested fix. A linter finding is reported with its triggered rule, location, and offending snippet, whereas an execution failure is reported with its error output and the relevant log excerpt. Both are accompanied by a suggested fix (Table~\ref{tab:build-repair-guidance}). The agent is required to read this feedback before editing the scripts, so that known failures are not repeated. Loop terminates if the construction has not succeeded after four attempts. The feedback follows the template shown in Figure~\ref{fig:build-feedback-template}.

\clearpage
\input{table/heatmap-box}

\input{table/build_fix_suggestion}
\begin{figure}[!htbp] 
    \centering
    
    \includegraphics[width=1.0\linewidth]{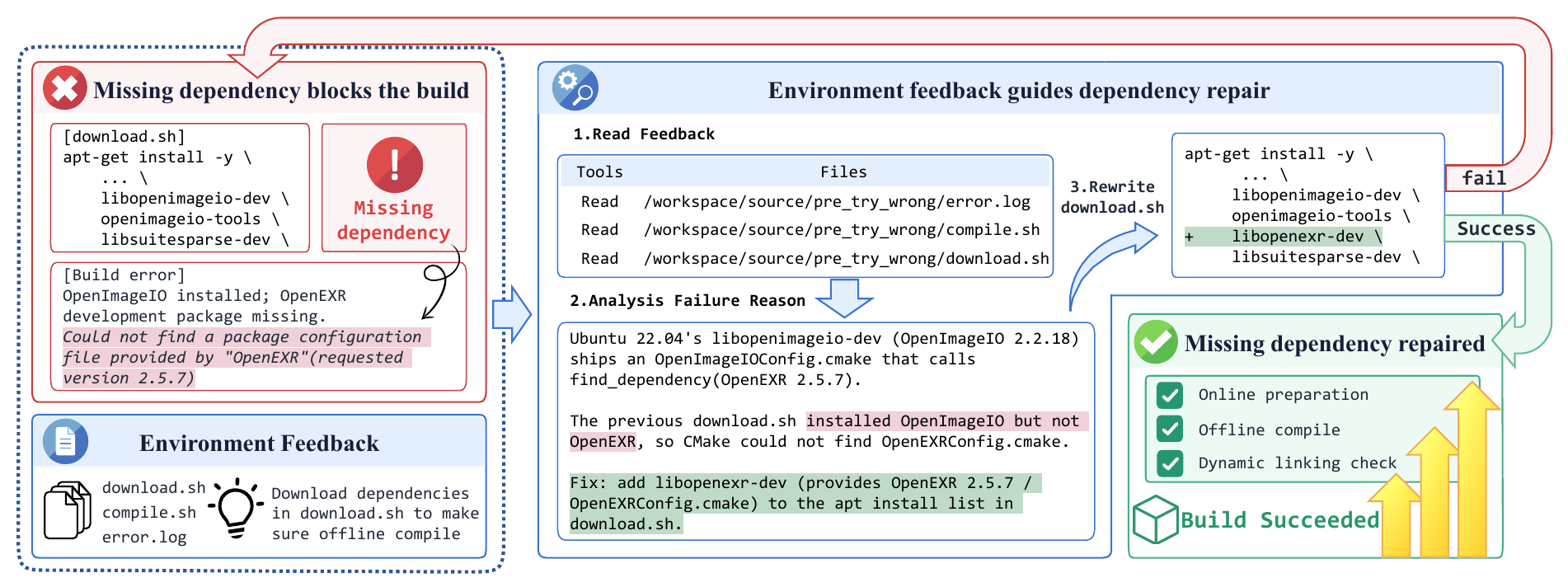} 
    \caption{A case of executable and sandbox generation guided by feedback. A missing package breaks offline compilation; the agent reads the feedback, adds \texttt{libopenexr-dev} to \texttt{download.sh}, and passes all checks on the next attempt.}
    \label{fig:build_retry}
\end{figure}

\paragraph{Case study.}
Figure~\ref{fig:build_retry} illustrates a repair guided by such feedback. In the first attempt, \texttt{download.sh} installs \texttt{libopenimageio-dev} but not the OpenEXR development package, so the offline build fails when CMake cannot locate \texttt{OpenEXRConfig.cmake}. The feedback packages the two scripts, the error log, and the remediation instruction to install the missing dependency in \texttt{download.sh}. The agent reads the log and scripts, attributes the failure to the absent \texttt{libopenexr-dev} package, and adds it to the \texttt{apt-get install} list. The revised scripts then pass online preparation, offline compilation, and the dynamic linking check.

\clearpage

\begin{figure}[t] 
    \centering
    
    \includegraphics[width=1.0\linewidth]{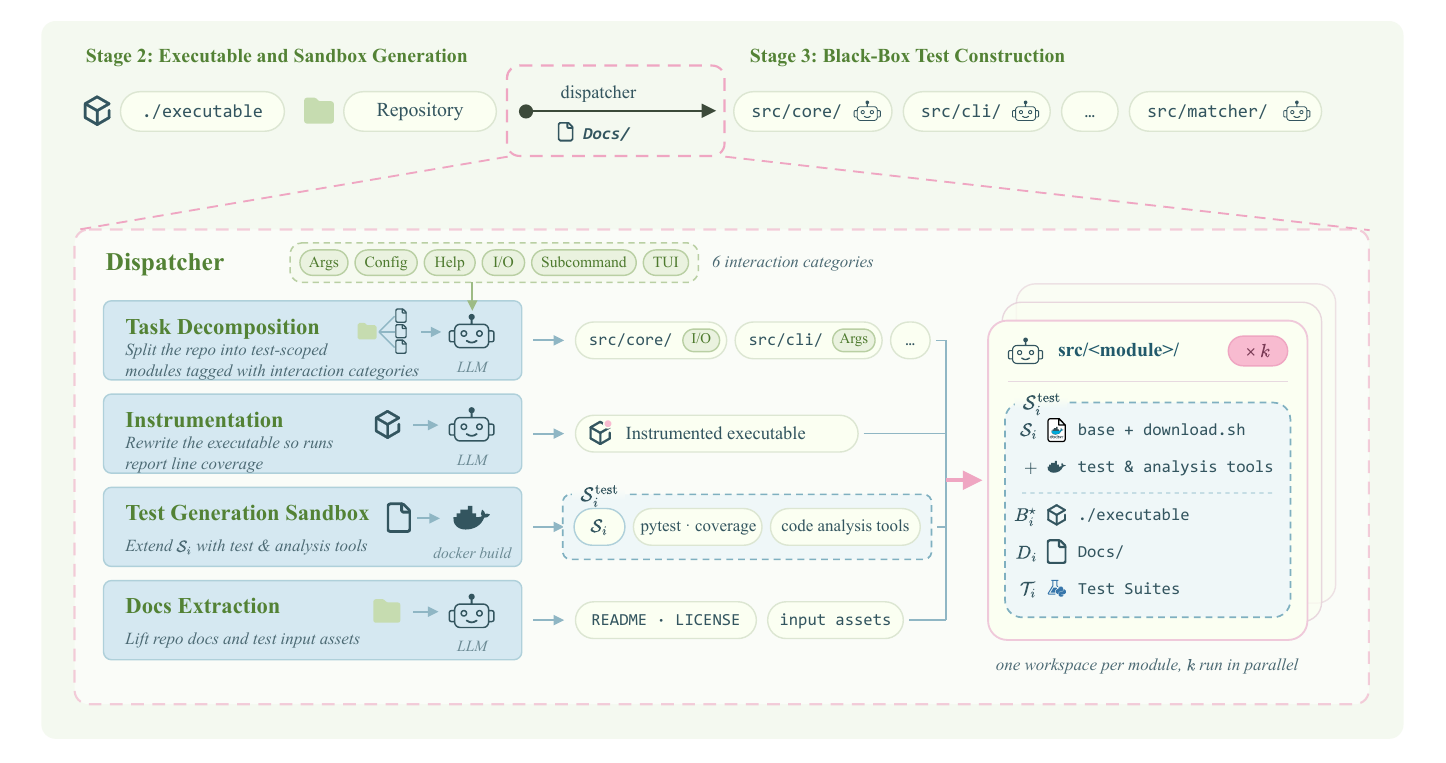} 
    \caption{Overview of the dispatcher bridging Stage~2 and Stage~3. It decomposes the repository into modules, instruments the executable for coverage, extends $\mathcal{S}_i$ into $\mathcal{S}_i^{\mathrm{test}}$, and extracts documentation $D_i$. Each module is then handed to a subagent in its own workspace, with $k$ workspaces running in parallel in Stage~3.}
    \label{fig:dispatcher}
\end{figure}


\subsection{Dispatcher Details}\label{Appendix:dispatcher}

As Figure~\ref{fig:dispatcher} shows, once the \emph{Executable and Sandbox Generation} completes, the pipeline invokes the dispatcher, which converts the validated target program into the resources required for test generation and evaluation. The dispatcher comprises four components: task decomposition, coverage instrumentation, test generation sandbox construction, and docs extraction. Instrumentation depends on the runtime sandbox $\mathcal{S}_i$, and sandbox construction depends on the instrumented program. The remaining components are independent and run concurrently, reducing the dispatcher's wall-clock time.

\subsubsection{Task Decomposition}
\label{appendix:decomposition}

Delegating an entire repository, often tens of thousands of lines, to a single agent rarely yields a test suite $\mathcal{T}_i$ with broad functional coverage, because the agent must simultaneously comprehend the codebase, prioritize among its functions, and remain within its context and generation budget. We therefore decompose the task. The dispatcher assigns a module budget according to repository complexity. Let $\ell$ denote the number of production-code lines and $r$ the number of runtime dependencies. We quantify complexity as
\begin{equation}
\xi = \operatorname{clip}_{[0,10]}\bigl(\log_{10}\ell + \log_{10}(1+r) - 2\bigr),
\end{equation}
where $\operatorname{clip}_{[0,10]}$ bounds the score to $[0,10]$. Repositories are classified as easy ($\xi<2$), medium ($2\le\xi<4$), or hard ($\xi\ge 4$), receiving module budgets of 4, 6, and 8, respectively. Larger budgets induce finer-grained decomposition, leading to more focused generation and higher-quality test suites.

A decomposition agent then partitions the repository under this budget into modules, each pairing a set of source paths with a functional description. The decomposition agent is instructed to cover the principal production source, keep each module functionally cohesive, and avoid file-level overlap across modules. The agent further tags each module with the applicable black-box interaction categories in Table~\ref{tab:6type}, and each subagent is prompted to target the behaviors in its tags. The distribution is shown in Table~\ref{tab:planning-input-categories}. If the repository ships a native test suite under \texttt{test/} or \texttt{tests/}, the dispatcher additionally appends a test-reuse module that rewrites the suite into the delivery contract without consuming the module budget. Each module is assigned to a dedicated subagent, and the $k$ subagents run in parallel to jointly compose $\mathcal{T}_i$. Parallelism reduces wall-clock generation time, while the narrowed scope of each subagent improves the quality of $\mathcal{T}_i$.

\input{table/6type}

\subsubsection{Coverage Instrumentation}
\label{appendix:instrumentation}

\paragraph{Overall strategy.}
To provide execution feedback to test generation, the dispatcher equips the target program with coverage collection, yielding the instrumented program. Rule-based instrumentation is the natural first choice, as it is deterministic and incurs no model cost. Build systems, however, vary considerably across repositories in their entry points, in how compiler flags are propagated, and in how source files map to build artifacts. A language-specific rule therefore captures the common instrumentation path of its language but not the build behavior peculiar to each repository, and even determining whether a rule applies to a given repository is nontrivial. Moreover, injecting instrumentation flags indiscriminately may corrupt the build by conflicting with existing compiler or linker settings (Table~\ref{tab:rule-instrumentation}, last column). We therefore adopt a \textbf{rule-first, agent-fallback} strategy. The language-specific rule first constructs an instrumentation candidate, which an independent validator then checks for the presence of build artifacts and coverage metadata. Only upon validation failure does the system fall back to an instrumentation agent, an LLM agent that diagnoses and repairs the instrumentation.

\paragraph{Language-specific rules.}
Table~\ref{tab:rule-instrumentation} summarizes the rule-based instrumentation method for each supported language (C, C++, Rust and Go), together with representative cases beyond rule-based instrumentation. These cases arise from repository-specific build behavior rather than from insufficiently detailed rules, so further refinement of the rules yields diminishing returns. Pairing language-specific rules with the instrumentation agent is thus the more effective combination.


\input{table/table_en}


\subsubsection{Test Generation Sandbox Construction}
\label{appendix:sandbox}

The subagents produced by task decomposition execute concurrently, which raises two issues. First, provisioning each subagent with its own copy of the program and repository incurs substantial space overhead. Second, concurrent writes to coverage counters introduce race conditions. We address both issues with two mechanisms.
\begin{enumerate}
  \item \textbf{Test generation sandbox.}  A repository-level sandbox for test generation $\mathcal{S}_i^{\mathrm{test}}$ is built on top of the runtime sandbox $\mathcal{S}_i$ produced by the Sandbox Generator, adding the language-specific coverage toolchain, and code analysis tools. Within $\mathcal{S}_i^{\mathrm{test}}$, a subagent analyzes the code, authors and executes tests, and reads back coverage entirely in place, without provisioning an environment of its own.
  \item \textbf{Shared artifacts with isolated state.} The validated instrumented program, runtime libraries, and coverage metadata are frozen as artifacts shared within a single pipeline attempt. For C/C++, these artifacts are baked into the image, and each execution instance receives its own copy-on-write layer for run state and coverage data, so that no two subagents write to the same counter file. Access is coordinated through leases and a concurrency budget, and results are written to per-consumer directories. Space overhead consequently drops from $O(k)$ to $O(1)$ in the number of subagents.
\end{enumerate}

\subsubsection{Docs Extraction}
\label{appendix:docs}

As shown in Figure~\ref{fig:dispatcher}, the dispatcher's fourth component, \emph{Docs Extraction}, employs an LLM agent to select two categories of content from the validated repository and place them into the documentation $D_i$ of the constructed environment. The first category is repository documents, namely the README and LICENSE. The second is test input assets that a model cannot reasonably synthesize on its own, such as images and domain-specific binary formats, together with the minimal runtime dependencies they require. The agent records its selection in a manifest, and the export copies only the listed files rather than auditing the whole repository. In addition, the pipeline rejects any source-code directory by rule. Together, these measures keep implementation details away from the solving side.

\subsection{Tests Construction Details}\label{Appendix:tests}

The construction of $\mathcal{T}_i$ follows an iterative process of test generation, validation, and refinement. In each round, subagents generate and improve tests for their assigned modules based on feedback from the environment. Section~\ref{appendix:test-validation} describes how generated tests are validated to ensure the reliability of $\mathcal{T}_i$. Section~\ref{appendix:test_feedback} then details how feedback is constructed to guide subsequent rounds until the final test suite is obtained.

\subsubsection{Validation for Test Construction}
\label{appendix:test-validation}


Validation for test construction ensures that the generated test suite $\mathcal{T}_i$ serves as a reliable criterion for the external behavior of the gold executable program $B_i^\star$. Concretely, $\mathcal{T}_i$ must be \textbf{faithful}, passing on the reference implementation, \textbf{stable}, with results that do not drift under execution order, \textbf{randomness}, and \textbf{effective}, constraining program functionality rather than passing unconditionally. Four mutually independent checks, run in parallel, enforce these conditions. The pass and double checks establish faithfulness and stability, while the dummy and weak-assert checks establish effectiveness dynamically and statically.

\paragraph{Pass check.}
Candidate tests are first executed on $B_i^\star$ to verify that their expected results, input conditions, and invocation are consistent with the reference behavior. The system records the outcome of each test case and marks those that fail or error on the reference implementation for revision. With $B_i^\star$ as the behavioral baseline, this check guarantees that $\mathcal{T}_i$ describes the actual behavior of the reference implementation.

\paragraph{Double check.}
A single pass on the reference implementation does not establish stability. The system therefore executes the same test set twice against the reference, isolating file state with separate working directories. The JUnit results of both runs are converted into normalized records containing the test identifier, its pass, fail, or skip status, and the corresponding diagnostic summary and output. If the two records disagree, the test is marked unstable and reported. This check prevents tests that pass by chance or depend on randomness from being accepted as stable criteria.

\paragraph{Dummy check.}
Passing stably on the reference program  $B_i^\star$ still does not show that a test constrains program behavior. The system therefore retains the same scripts and inputs but replaces the program under test with a dummy implementation that produces no meaningful output, and executes the suite again. A test that still passes is marked \texttt{dummy\_pass}. This check identifies tests that do not truly depend on the functionality under test, such as those asserting only process termination or tautological conditions.

\paragraph{Weak-assert check.}
Runtime controls expose only extreme cases and say little about assertion strength. The system therefore audits each test with a static linter of 37 rules (Tables~\ref{tab:linter-rule-inventory-1} and~\ref{tab:linter-rule-inventory-2}), which rejects weak oracles such as sole exit-code checks or tautological assertions, assertions masked by conditionals or swallowed exceptions, access to the reference source, and nondeterministic dependencies such as network, time, or unseeded randomness. Findings are reported at two levels. Errors mark defective tests that are excluded from the final suite, whereas warnings flag risks to be reviewed against task semantics, such as floating-point precision or unordered output, where implementation differences may be permissible.

Formally, let $\tilde{\mathcal{T}}_i$ denote the candidate suite, $B_\varnothing$ the dummy implementation, and $t(B)\in\{\textsc{pass},\textsc{fail}\}$ the outcome of test $t$ on program $B$, with $t^{(1)}$ and $t^{(2)}$ denoting two independent runs. Let $\mathcal{L}(t)$ be the set of error-level findings reported by the linter. A test is accepted into $\mathcal{T}_i$ only if it satisfies all four checks:
\begin{equation}
\mathcal{T}_i=\Bigl\{\,t\in\tilde{\mathcal{T}}_i \;\Bigm|\;
\underbrace{t(B_i^\star)=\textsc{pass}}_{\text{pass check}}\;\wedge\;
\underbrace{t^{(1)}(B_i^\star)=t^{(2)}(B_i^\star)}_{\text{double check}}\;\wedge\;
\underbrace{t(B_\varnothing)=\textsc{fail}}_{\text{dummy check}}\;\wedge\;
\underbrace{\mathcal{L}(t)=\varnothing}_{\text{weak-assert check}}
\,\Bigr\}.
\label{eq:test-validation}
\end{equation}

The rejected tests $\tilde{\mathcal{T}}_i\setminus\mathcal{T}_i$, together with all warnings, are consolidated into a structured quality report. The report is written to \texttt{feedback\_docs/} and returned to the next iteration as environment feedback for revising $\tilde{\mathcal{T}}_i\setminus\mathcal{T}_i$, as detailed in Appendix~\ref{appendix:test_feedback}.

\input{table/linter}

\subsubsection{Feedback for Test Generation}
\label{appendix:test_feedback}
The raw observations available after a round, namely execution logs, validation reports, and coverage data, have low information density. To guide the agent effectively, the pipeline processes them into structured feedback that answers three questions: which existing tests need repair, which uncovered code the next round should target, and through which public entries that code can be reached. Tests to repair are identified by the four checks, whereas new tests are guided by program analysis. Each round thus repairs first and then writes new tests. Under the guidance of feedback, effective and deterministic tests are written without blind exploration.

\paragraph{Validation feedback and repair.}
After each round, the four checks of Section~\ref{appendix:test-validation} write their findings to \texttt{validation.json}. Rather than raw pytest output, each finding is reported as an actionable item with the test or script location, issue category, severity, diagnostics, and a suggested fix, so that the agent knows not only what failed but also how to correct it. Repair precedes authoring for two reasons. First, repairing an existing test is far cheaper than writing a new one, since the location and fix are already identified and the change is typically local. Second, newly written tests may introduce further issues, and without timely repair these defects accumulate across rounds. Repairing first keeps every retained test stable and effective, so that coverage grows on a reliable foundation.

\paragraph{Writing guided by dynamic program  analysis.}
A trustworthy verifier must cover as much functionality as possible, so the central question in writing new tests is which functionality remains unreached. Software engineering answers this with coverage, since unexecuted source locations largely correspond to unexplored functionality. Each language uses a coverage backend matched to its instrumented program, as shown in Table~\ref{tab:test-feedback-backends}. All backends report file path, valid counting units, covered units, and uncovered locations. C/C++ and Rust count executable source lines, whereas Go retains its native statement-weighted basic blocks.

\input{table/backends}

To make coverage reflect core functionality rather than dependencies, tests, or build artifacts, the system filters core files by source-ownership rules and configured path scopes. Let $\mathcal{F}_{\mathrm{core}}$ denote the set of core files, $\mathcal{U}_f$ the valid counting units of file $f$, $w_u$ the weight of unit $u$, and $h_t(u)\in\{0,1\}$ whether $u$ is hit in round $t$. Core coverage is defined as
\begin{equation}
 C_{\mathrm{core}}^{(t)} =
 \frac{\sum_{f\in\mathcal{F}_{\mathrm{core}}}\sum_{u\in\mathcal{U}_f} w_u h_t(u)}
      {\sum_{f\in\mathcal{F}_{\mathrm{core}}}\sum_{u\in\mathcal{U}_f} w_u}.
 \label{eq:test-core-coverage}
\end{equation}
For C/C++ and Rust, each unit $u$ is an executable source line with $w_u=1$. For Go, each unit is a basic block weighted by its number of statements. Summing over units rather than averaging per-file coverage prevents small files from carrying the same weight as large ones. The denominator counts only source units the backend can attribute to a file, not physical lines. In addition, a report with no such units or a failed collection is discarded rather than treated as zero coverage.


A raw coverage report often runs to tens or even hundreds of megabytes, most of it recording lines that are already covered or belong to files outside the module. The full report is therefore kept only for auditing, and the test generation agent receives a processed \texttt{coverage\_targets.json}. Let $\mathcal{E}_t$ be the set of core files measured in round $t$ and $c_t(f)\in[0,1]$ the coverage of file $f$. The partially covered files $\mathcal{P}_t$ and the target set $\mathcal{G}_t$ are defined as
\begin{equation}
 \mathcal{P}_t=\{f\in\mathcal{E}_t:0<c_t(f)<1\},\qquad
 \mathcal{G}_t=
 \begin{cases}
   \mathcal{P}_t, & \mathcal{P}_t\ne\varnothing,\\
   \{f\in\mathcal{E}_t:c_t(f)<1\}, & \text{otherwise}.
 \end{cases}
 \label{eq:test-feedback-targets}
\end{equation}
That is, $\mathcal{G}_t$ consists of partially covered files if any partially covered file exists, and falls back to all uncovered files otherwise.
Partially covered files $\mathcal{P}_t$ are selected first. Their existing coverage shows that the current tests already reach them, so the next round only needs to exercise the remaining branches and inputs. Fully covered files are excluded since they offer nothing more to gain. Files with zero coverage are deferred. Without execution evidence, the agent cannot tell whether such a file holds untriggered functionality, depends on network access, or is simply unreachable through the black-box interface of the executable, so exploring it would be a blind search. They are considered only when no partially covered file $\mathcal{P}_t$ remains.

Candidates are sorted by ascending coverage so that the files farthest from completion are presented first. The target set is often large, and an agent handed all of it at once tends to overclaim, reporting targets as covered without actually testing them. Each round therefore provides at most $\min(|\mathcal{G}_t|,\lceil\rho|\mathcal{G}_t|\rceil,B)$ targets, with feedback ratio $\rho=0.1$ and cap $B=300$, so that the agent works through a manageable subset in each round. The selected targets, each with its path, coverage, unit counts, and uncovered source lines, are written to \texttt{coverage\_targets.json}. A companion \texttt{coverage\_summary.md} reports core coverage and its change, the measurement scope, and the measurement status. 

\paragraph{Writing guided by static program analysis.}
An uncovered location alone does not indicate which input to construct, particularly when the target function sits deep in a call hierarchy. Program-Env builds on mature static analysis tools rather than custom analyzers. Joern provides code property graphs (CPG)~\citep{yamaguchi2014joern}, call graphs, control flow, and data flow for all supported languages except Rust, for which CodeQL is used instead. Because these tools expose standard interfaces, the analysis layer is pluggable, and a backend can be replaced or a new analysis added without changing the rest of the pipeline. The databases produced by Joern and CodeQL span the entire program and are far too large to hand to the agent. Program-Env therefore queries them only for the uncovered targets in \texttt{coverage\_targets.json}, which is why static analysis starts from the second round, after the first coverage measurement. For each target, the pipeline maps its uncovered lines to the enclosing functions, traverses the call graph in reverse to find public entries through which a black-box test can reach them, and gathers the control-flow and data-flow context of those functions. The results are written to \texttt{feedback\_docs/}. Beyond the pre-computed artifacts, the Joern and CodeQL databases remain queryable inside $\mathcal{S}_i^{\mathrm{test}}$, and the agent is encouraged to issue its own queries while writing tests.
\paragraph{Cross-round summary.}
The feedback above describes only the current round, while the agent's context is not preserved across rounds. To serve as memory between rounds, the agent writes \texttt{summary.md} at the end of each round, recording what it did, which targets it addressed or skipped and why, and which targets are left for the next round. The next round reads this summary before acting, so that it continues from the previous state rather than starting from scratch.


\input{table/summary}


\paragraph{Feedback layout.}
All feedback processed by Program-Env resides under \texttt{Feedback\_Docs/}, organized by module and agent role as shown below.
A skill we prepare in advance prescribes the intended workflow for test writing, including the order in which the agent reads these documents: \texttt{module\_assignment.json} for its assigned scope, \texttt{summary.md} for what the previous round accomplished, \texttt{validation.json} for how to repair the flagged tests, and \texttt{coverage\_targets.json} for low-coverage files, which it combines with the call graph and control-flow records in \texttt{static\_analysis\_result/} to locate public entries and write new tests.
\input{table/feedback_file_tree}

\begin{figure}[h] 
    \centering
    
    \includegraphics[width=1.0\linewidth]{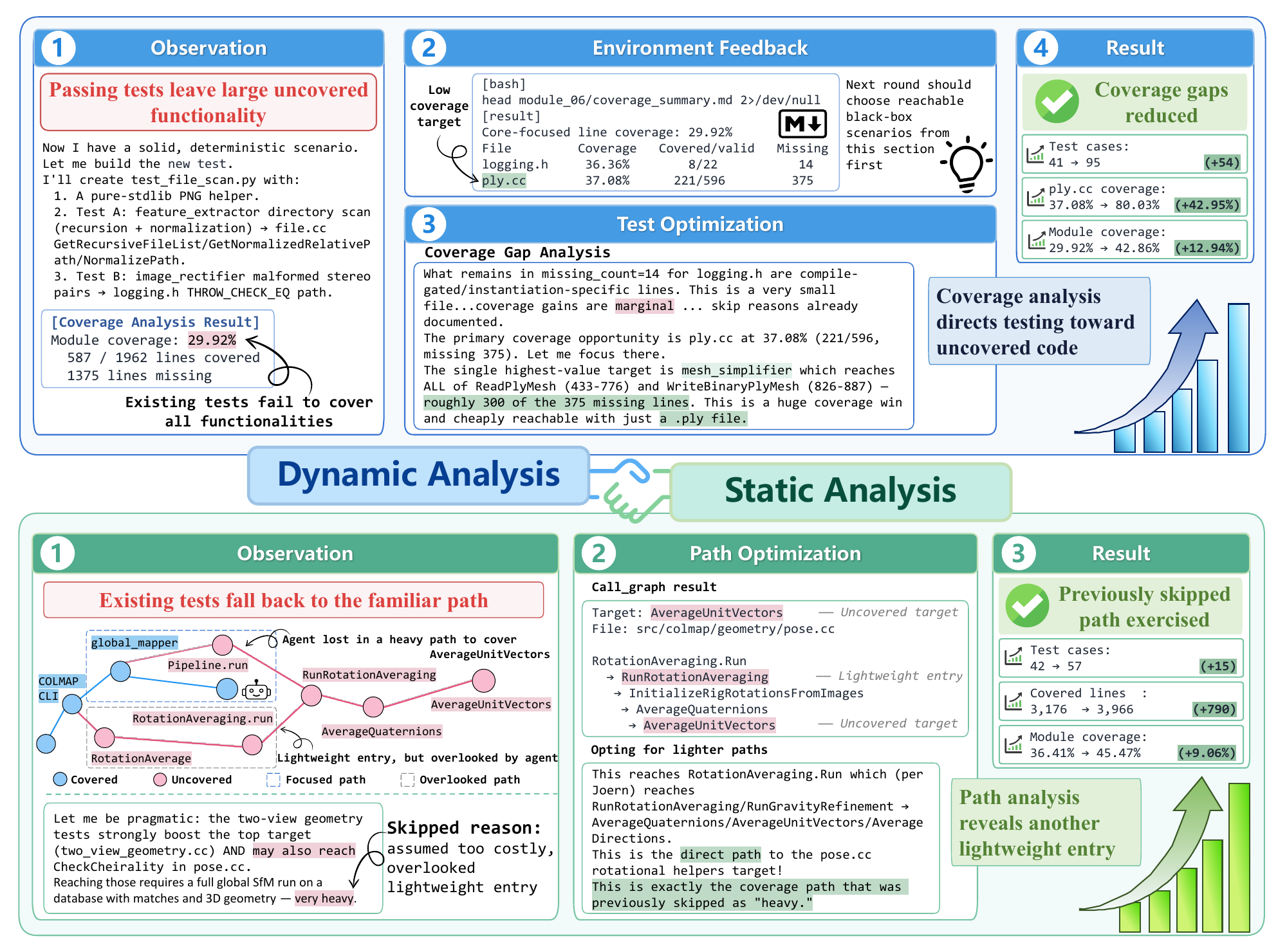} 
    \caption{Program analysis guiding test writing on COLMAP. Top: coverage feedback (dynamic) steers the agent toward uncovered code in \texttt{ply.cc}. Bottom: the call graph (static analysis) reveals a lightweight entry to a target the agent had skipped.}
    \label{fig:test_retry}
\end{figure}

\paragraph{Case study.}
Figure~\ref{fig:test_retry} illustrates both kinds of guidance on COLMAP. The first case shows dynamic analysis. All tests pass yet module coverage is only 29.92\%. Coverage feedback singles out \texttt{ply.cc} with 375 missing lines, and the agent finds that one \texttt{.ply} input through \texttt{mesh\_simplifier} reaches most of them, lifting the file to 80.03\% and the module to 42.86\%. The second case shows static analysis. The agent had skipped \texttt{AverageUnitVectors}, assuming it was reachable only through a full structure-from-motion run and dismissing that path as too heavy. The call graph reveals a lightweight path through the standalone \texttt{rotation\_averaging} command that reaches the target left uncovered in the previous round, and following it raises module coverage from 36.41\% to 45.47\%. In both cases, feedback turns a guess about what to test next into a clear, reachable target.

\begin{figure*}[t]
    \centering
    \begin{subfigure}[t]{0.49\textwidth}
        \centering
        \includegraphics[width=\linewidth]
            {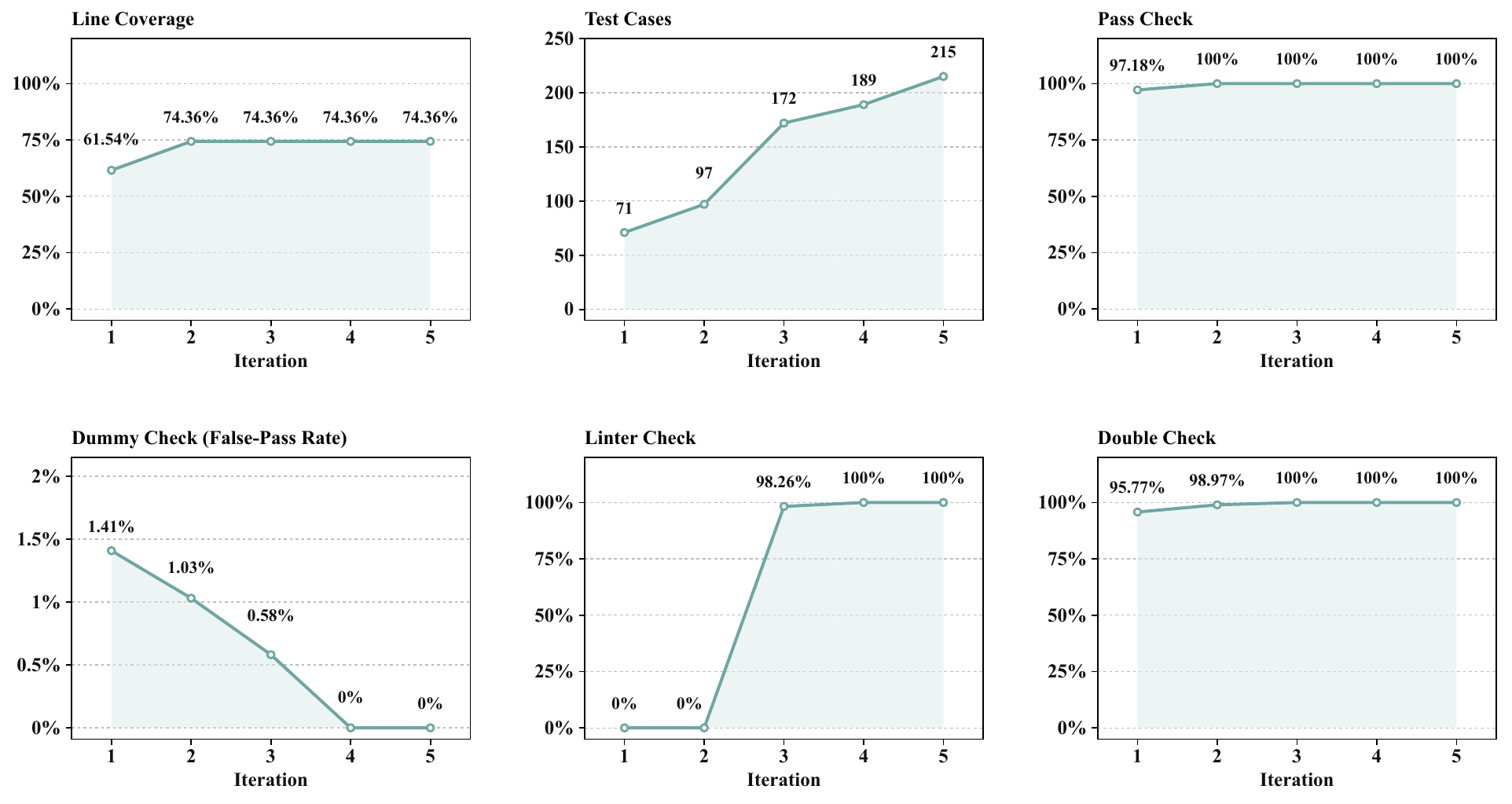}
        \caption{C: \texttt{paullouisageneau/violet}.}
        \label{fig:feedback-violet}
    \end{subfigure}
    \hfill
    \begin{subfigure}[t]{0.49\textwidth}
        \centering
        \includegraphics[width=\linewidth]
            {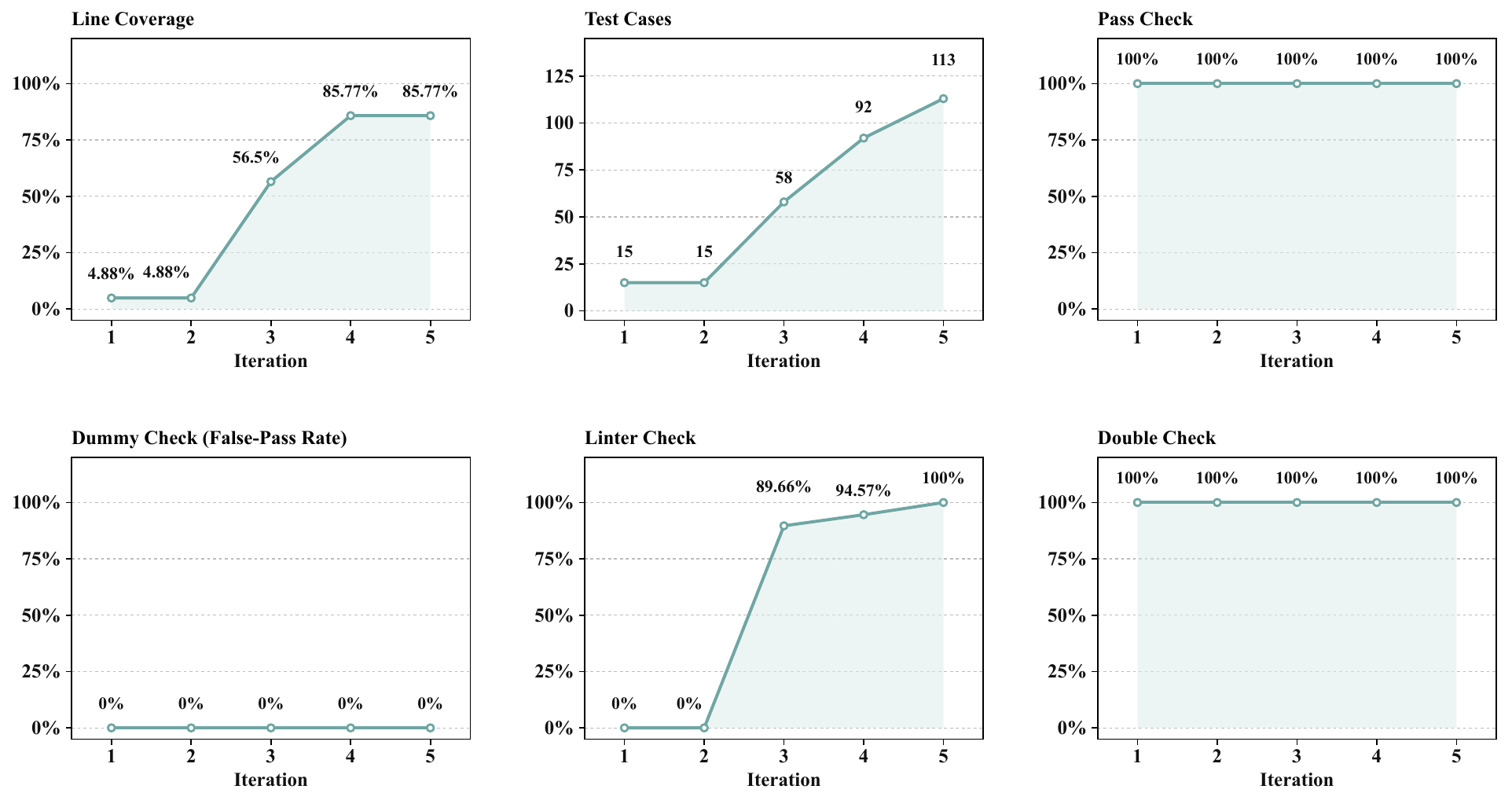}
        \caption{Rust: \texttt{Ramilito/kubesess}.}
        \label{fig:feedback-kubesess}
    \end{subfigure}

    \caption{Iteration-wise coverage, test counts, and validation
    metrics for two representative source modules. \textbf{(a)} \texttt{paullouisageneau/violet} shows the agentic loop in Program-Env repairing weak assertions and dummy false passes.\textbf{ (b) }\texttt{Ramilito/kubesess} shows it steadily raising coverage under feedback.}
    \label{fig:feedback-cases}
\end{figure*}

Figure~\ref{fig:feedback-cases} shows the effect of feedback across iterations. In \texttt{paullouisageneau/violet}, feedback guides the agent to repair weak assertions and dummy false passes by the 4th round, raising the linter pass rate from 0\% to 98.3\% and reducing the false-pass rate from 1.41\% to 0\%. In \texttt{Ramilito/kubesess}, coverage feedback raises coverage from 4.88\% to 85.77\%. 

\subsubsection{Loop termination}\label{appendix:loop-termination}

A subagent terminates once its module reaches the functionality coverage threshold or the round limit. The round limit scales with repository complexity $\xi$ as $R=\operatorname{clip}_{[5,20]}\bigl(5\lfloor\xi\rfloor\bigr)$, granting harder repositories more rounds. Some targets, however, are unreachable through black-box testing, and continuing to pursue them wastes budget without improving $\mathcal{T}_i$. We therefore add an early stop. It compares round $t$ with round $t-3$ and terminates the subagent when core coverage has gained less than 0.25\% and the linter pass rate has gained less than 20\%. The earliest possible stop is at round 5. Table~\ref{tab:test-generation-stopping} summarizes these settings.
\input{table/early_stop}

Once all subagents terminate, the system runs a final validation over the test branches they have written. Tests that fail any of the four checks are marked as ignored with the reason recorded, and the remaining test scripts, required resources, test lists, and ignore records from all branches are exported together as the delivered test suite $\mathcal{T}_i$.

\subsection{Additional Dataset Statistics}
\label{appendix:Statistics}

This section characterizes Program-Env along five dimensions, namely community activity, repository scale, functional category, construction cost, and verification quality, showing that the dataset is both high in quality and broad in diversity.

\begin{figure}[h] 
    \centering
    
    \includegraphics[width=0.9\linewidth]{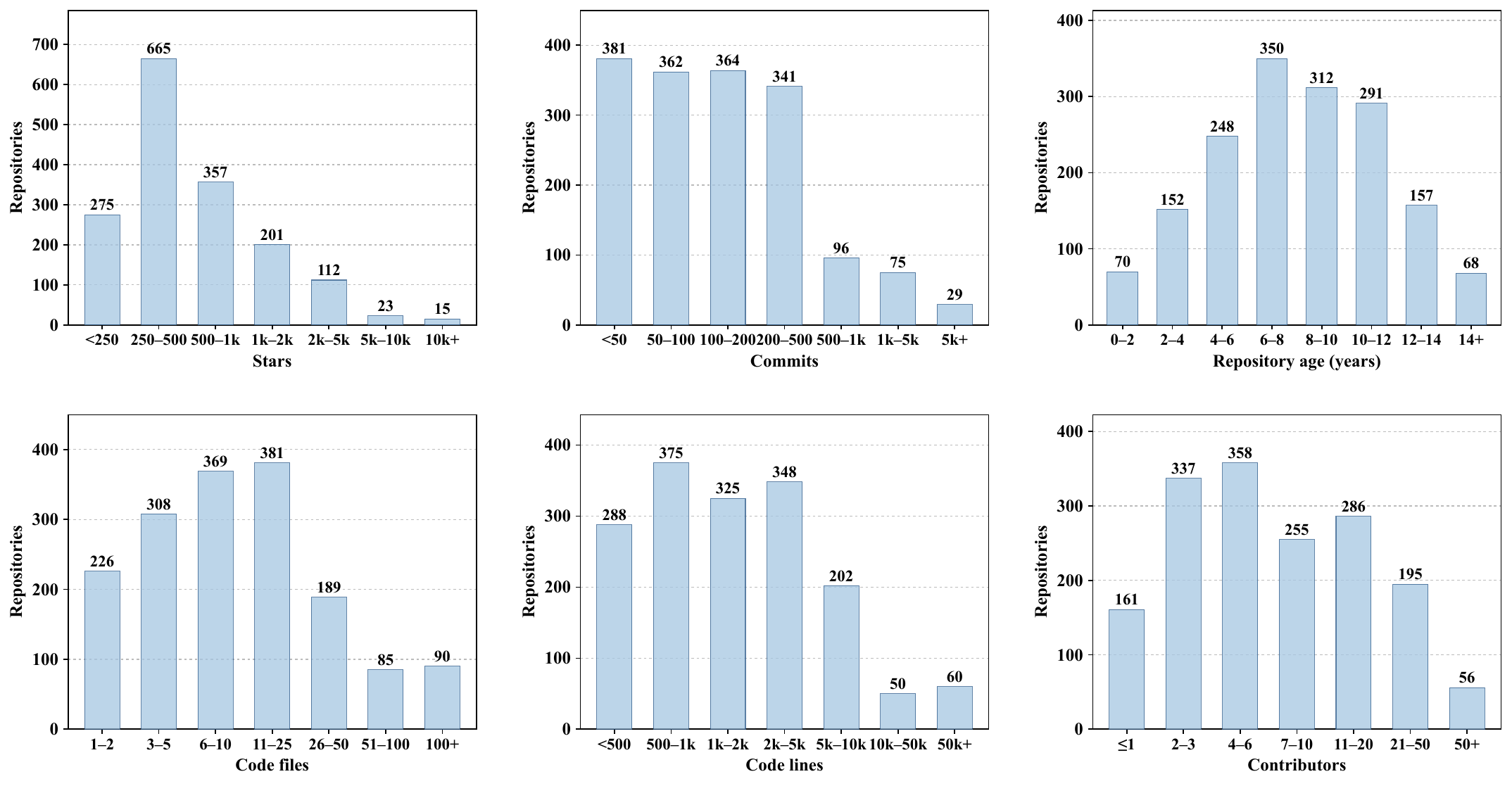} 
    \caption{Distributions of community activity, development history, and codebase size across the 1,648 collected repositories.}
    \label{fig:image_dataset_overview}
\end{figure}

\begin{figure}[h] 
    \centering
    
    \includegraphics[width=0.5\linewidth]{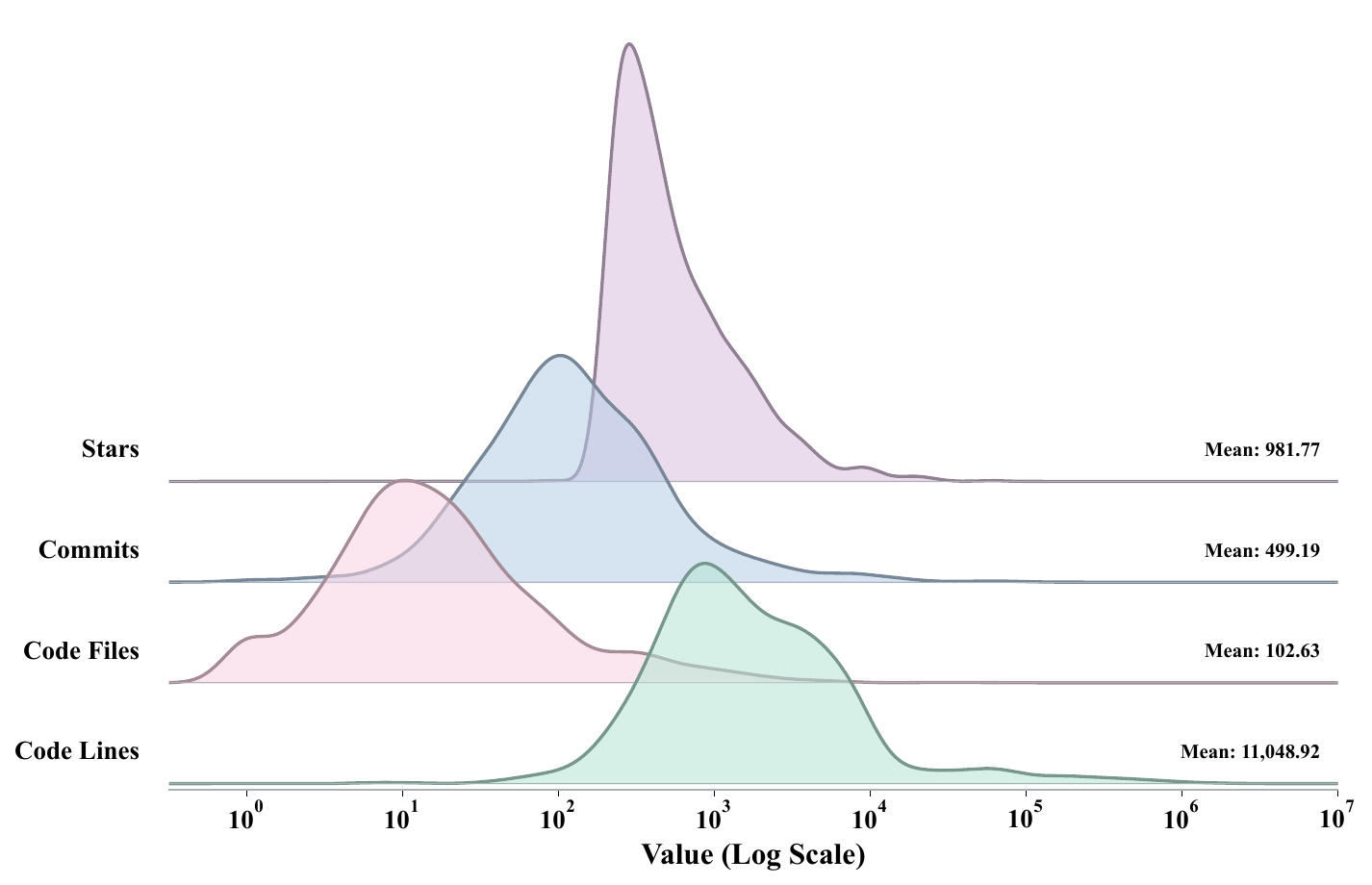} 
    \caption{Kernel density estimates of stars, commits, code files, and code lines across the 1,648 collected repositories, on a log scale. Means are annotated on the right.}
    \label{fig:density_repo}
\end{figure}

\paragraph{Community activity.}
Figure~\ref{fig:image_dataset_overview} and Figure~\ref{fig:density_repo} show the distributions of stars, commits, repository age, and contributors across the 1,648 repositories. Stars and commits are both right-skewed with long tails, with means of 981.77 and 499.19 and medians of 425 and 113. The dataset therefore spans mature, widely adopted projects with long collaborative histories as well as small domain-specific tools, covering a wide range of attention, collaboration scale, and development history.

\paragraph{Repository scale.}
As shown in Figure~\ref{fig:image_dataset_overview}, production code lines and code files average 11,048.92 and 102.63 per repository and range from hundreds of lines to over one million. Most repositories are of moderate size, but the dataset also retains both much smaller and much larger projects, so the environments cover a wide range of scale.

\paragraph{Functional categories.}
We group repositories into eight functional categories by their primary purpose. The categories are derived from the Debian package sections, merged into eight labels that fit the collected repositories. Each repository is assigned by inspecting its README, documentation, and source code, and multi-purpose projects are placed in a single dominant category. Table~\ref{tab:functional-categories} gives the definitions and counts, and Figure~\ref{fig:image_dataset_overview_repo} shows the resulting proportions.

\input{table/8type}

\clearpage

\begin{figure}[h] 
    \centering
    
    \includegraphics[width=0.45\linewidth]{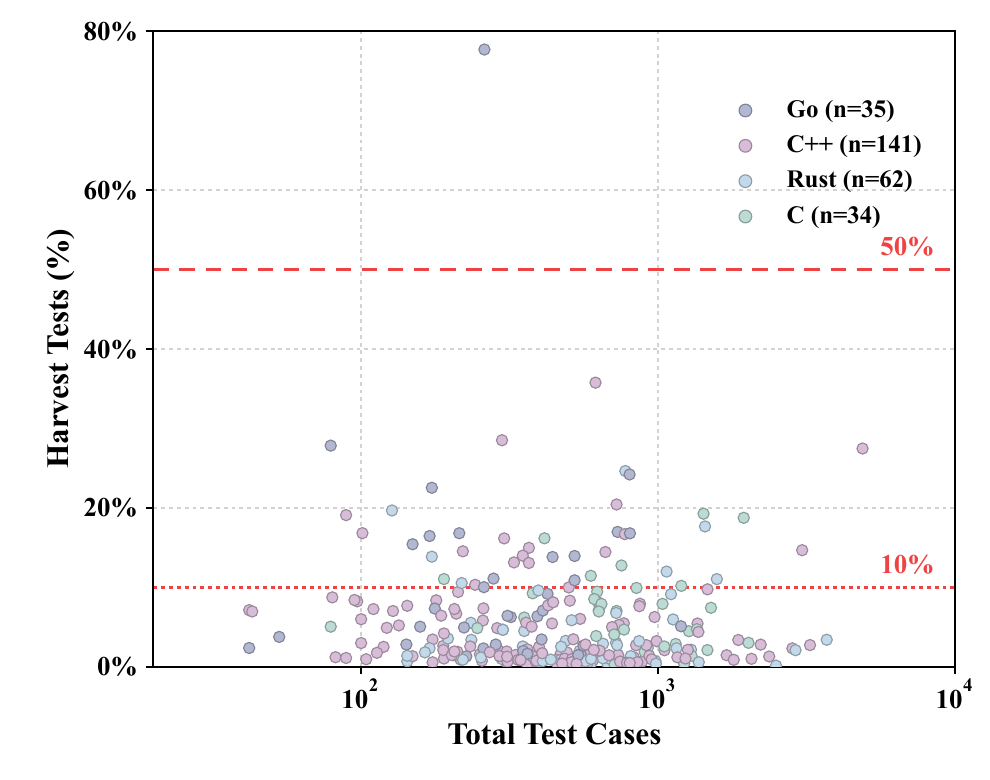} 
    \caption{Tests harvested from existing repository test suites, for the 272 repositories that ship native tests. Each point is one repository, plotted against its total test count.}
    \label{fig:harvest_scatter}
\end{figure}

\begin{figure}[h] 
    \centering
    
    \includegraphics[width=.9\linewidth]{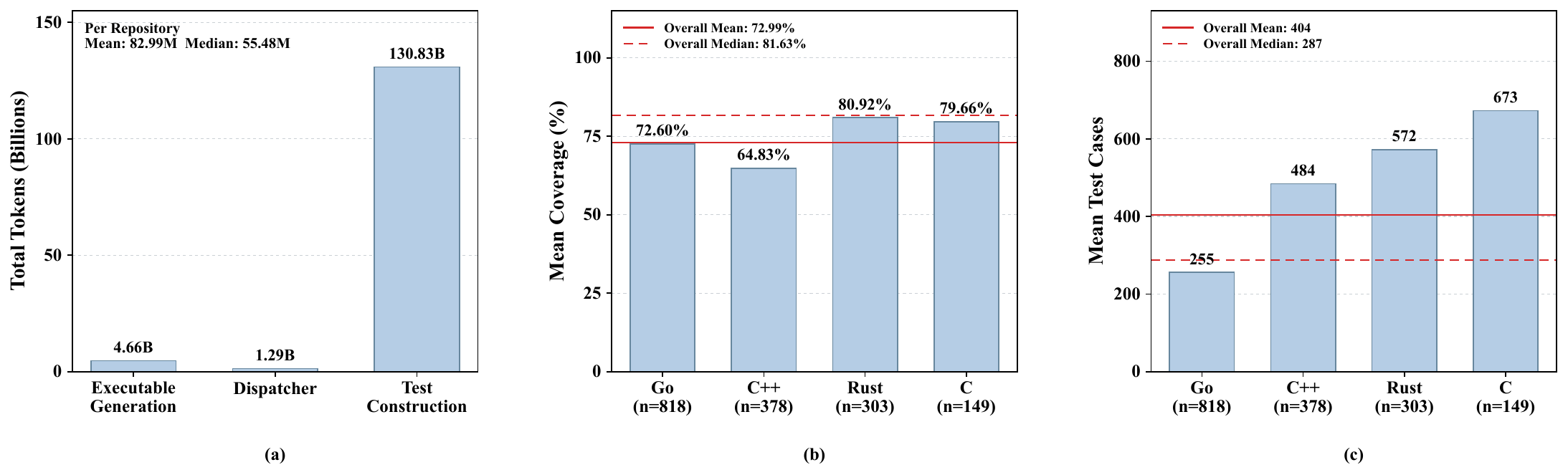} 
    \caption{Token cost and generation quality of Program-Env.  Dispatcher stage covers the LLM calls for coverage instrumentation, task decomposition and docs extraction.
    }
    \label{fig:dataset_overview_token}
\end{figure}

\paragraph{Construction cost.}
As shown in Figure~\ref{fig:dataset_overview_token}, building the 1,648 tasks consumed 136.77B tokens in total, the vast majority on writing tests. Of the final 665,512 test cases, 655,559 (98.50\%) are written from scratch by the pipeline against source functionality, and only 9,953 (1.50\%) are rewritten from the repository's native tests. Even among the 272 repositories that ship a native test suite, reused tests account for a median of 2.8\% of the final suite (Figure~\ref{fig:harvest_scatter}), so the verifiability of each environment comes from the pipeline rather than from upstream tests. Static program analysis costs 11.74 hours in total (Table~\ref{tab:static-analysis-cost}), yet scanning a single repository takes only 25.64 seconds on average.
\begin{table}[h]
\centering
\small
\caption{Cost of static program analysis.}
\renewcommand{\arraystretch}{1.15}
\begin{tabular}{@{}lrrrr@{}}
\toprule
\textbf{Tool} & \textbf{Repositories} & \textbf{Total time} & \textbf{Avg. per repo}  \\
\midrule
Joern  & 1,345 & 7.00 h & 18.73 s  \\
CodeQL &   303 & 4.74 h & 56.29 s  \\
\midrule
Total  & 1,648 & 11.74 h & 25.64 s \\
\bottomrule
\end{tabular}
\label{tab:static-analysis-cost}
\end{table}

\clearpage

\begin{figure}[h] 
    \centering
    
    \includegraphics[width=0.85\linewidth]{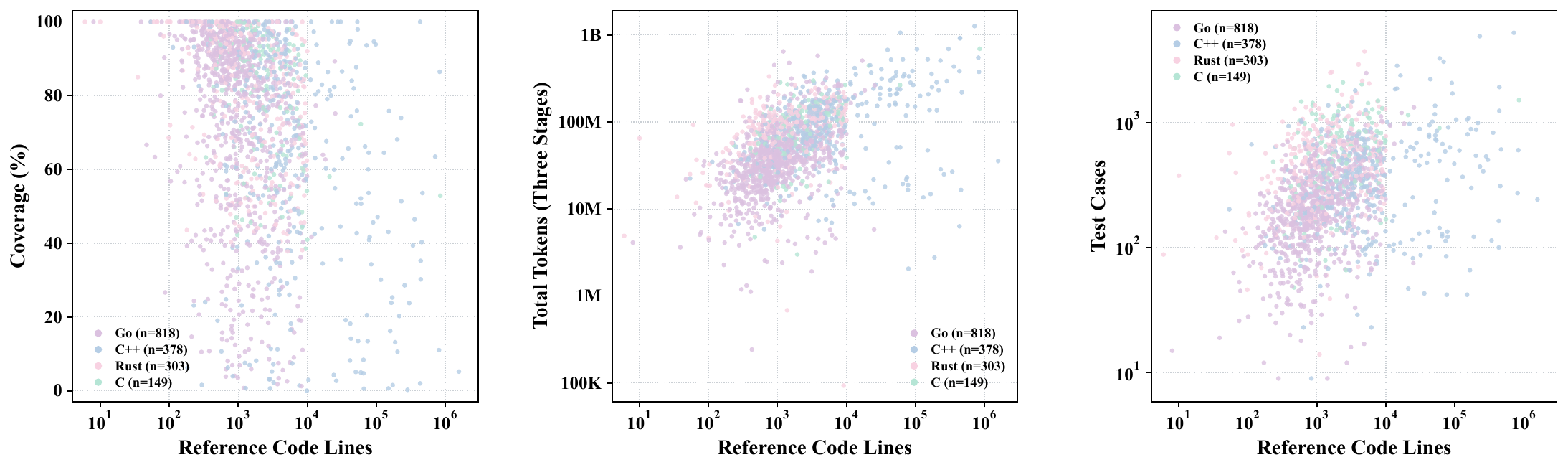} 
    \caption{Coverage, total tokens across the three stages, and test cases as a function of reference code lines. Each point is one repository, colored by language. Token and test counts grow with repository size.}
    \label{fig:scatter}
\end{figure}

\paragraph{Verification quality.}
As shown in Figure~\ref{fig:dataset_overview_token}, the final dataset reaches a mean coverage of 72.99\% (median 81.63\%) and contains 665,512 test cases. Figure~\ref{fig:scatter} further shows that test count and token consumption grow with repository size while coverage stays high even for the largest and hardest repositories. Full coverage is not the goal, since not every function of a repository is reachable through its black-box interface. As long as $\mathcal{T}_i$ faithfully captures the reachable behavior of $B_i^\star$, verification quality is unaffected.

\clearpage
\section{Program-Agent Details}\label{app:program-agent}

\subsection{Phase Orchestration}
\label{app:program-agent-orchestration}

Program-Agent coordinates its analysis, implementation, and review stages
through a fixed-round external orchestrator. Given a budget of $K$
implementation rounds, it executes one initial analysis, $K$ implementation
stages, and $K-1$ review stages. No review is performed after the final
implementation because review artifacts are used exclusively to guide the next
implementation round.

Each stage invocation runs in a fresh model session. Instead of propagating the
complete conversational history, the orchestrator transfers persistent
information through Handoff artifacts. The workflow follows the
predefined revision budget and does not parse a stopping decision from the
review output.

Let $D$ denote the task instruction and project documentation,
$B^\star$ the execute-only reference program, and $W$ the source workspace
shared across implementation rounds. We use $A$ for the initial analysis
artifact, $I_k$ for the delivery artifact of implementation round $k$, $R_k$
for the corresponding review artifact, and $C_k$ for the committed source
revision produced in round $k$. The executable $E_k$ is independently built
from $C_k$ for validation and differential review; it is a derived artifact
rather than the primary task deliverable. For $k=1$, the sequence of preceding
review artifacts $R_1,\ldots,R_{k-1}$ is empty.

\input{table/pb-agent_pseudo-code}





\subsection{Workspace and Access Isolation}
\label{app:program-agent-isolation}
Program-Agent separates product source code, inter-stage artifacts, and private orchestration state. The shared Workspace stores the source implementation that evolves across implementation rounds, Handoff stores the persistent artifacts exchanged between model sessions, and a private workflow directory stores build snapshots, failure diagnostics, and archived session state. Their logical layout is shown below.

\input{table/pb-agent_layout}

In the implementation, \texttt{/handoff} is a symbolic link to artifact storage managed by Harbor~\citep{Harbor_Framework}, allowing the analysis, implementation, and review artifacts to be inspected while the workflow is running. In contrast, the private workflow directory is accessible only to the orchestrator during model execution. Its build snapshots, failure diagnostics, session records, and workflow state are exported to \texttt{/logs/agent/workflow} after all model stages terminate. The Revisions region in Figure~\ref{fig:program-agent} corresponds logically to the validated snapshots stored under \texttt{workflow/builds/rev-k}.

The three stages operate on different workspace views and receive different forms of access to the reference program, candidate program, product source, Handoff artifacts, and private orchestration state. Their resource access is summarized in Table~\ref{tab:program-agent-access}.

\input{table/pb-agent_access}

Analysis and Review operate on disposable source copies and are prohibited by their role contracts from modifying product source, whereas Implement develops directly in the shared Workspace. This role-level restriction differs from the filesystem-enforced isolation applied to the reference executable, candidate snapshots, and Handoff artifacts.

\paragraph{Analysis isolation.}
Analysis runs in a disposable copy of the initial Workspace. The executable originally present in the Workspace is excluded from this copy and replaced with an execute-only reference program. The session is instructed not to modify product source and may write persistent output only to its active Handoff directory. Once Analysis is complete, its disposable workspace is removed.

\paragraph{Implementation isolation.}
Before the first Implement session starts, the reference executable is removed from all implementation-visible paths and retained only in root-owned private storage. Implement can therefore access the frozen analysis and review artifacts together with its own candidate, but cannot execute or inspect the reference program. It is the only stage that directly modifies product source in the shared Workspace.

\paragraph{Review isolation.}
Each Review session runs in a separate disposable workspace. This workspace contains a snapshot of the current source tree, the initial analysis, all preceding reviews, the current implementation artifact, an execute-only copy of the reference program, and an immutable copy of the independently built candidate. The role may execute both programs through their documented interfaces but is instructed not to modify product source, build scripts, or version history. The entire review workspace is deleted after its artifact has been sealed.

\paragraph{Artifact immutability.}
The Handoff root and all completed stage directories remain read-only during model execution. Only the subdirectory owned by the active stage is writable. After successful stage validation, the orchestrator transfers that directory to root ownership and recursively removes its write permissions. Downstream sessions may therefore consume completed artifacts but cannot alter the evidence, implementation reports, or requirements recorded by earlier stages.

The \texttt{/logs/agent} directory is also read-only during model execution, preventing sessions from creating undeclared persistent state outside Handoff. Analysis and Review may create temporary fixtures in their disposable workspaces, but these fixtures are deleted with the workspace unless they are explicitly retained and documented in the active Handoff artifact.

\subsection{Build Validation and Failure Recovery}\label{app:program-agent-recovery}
A stage is accepted only after both the model invocation and its stage-specific machine checks succeed. Analysis and Review must produce their required non-empty reports, whereas Implement must additionally satisfy source-state and independent-build requirements. These checks are summarized in Table~\ref{tab:program-agent-validation}.

\input{table/pb-agent_validation}

\paragraph{Source-state validation.}
After an Implement session terminates, the orchestrator verifies that \texttt{delivery.md} is non-empty, no tracked file has staged or unstaged changes, and \texttt{compile.sh} is tracked by Git. The current Git \texttt{HEAD} is then recorded as the source revision for the round. Untracked files do not directly fail this check, although the role contract requires all intended source files to be committed and generated artifacts to be ignored.

\paragraph{Independent build.}
The orchestrator exports a clean Git archive of the validated source revision into a new build directory rather than compiling directly in the development Workspace. It removes common HTTP proxy, Python package index, and npm registry variables before running \texttt{compile.sh} with a 300-second timeout. A build is successful only if the script terminates successfully and produces a file named \texttt{executable}.

For each successful build, Program-Agent records the source commit, archived candidate path, and SHA-256 digest in \texttt{build.json}. The same record is copied to the current implementation artifact as \texttt{independent-build.json}. The resulting executable is then used as the immutable candidate for Review, while the committed source tree and \texttt{compile.sh} remain the primary task deliverable.

\paragraph{Bounded retry.}
Each stage has an independent attempt budget, configurable from one to three attempts and set to two by default. Every retry starts a fresh model session. The new session receives only a bounded diagnostic describing the previous delivery or build failure, rather than the previous transcript.

When Analysis or Review fails validation, its stage directory is deleted before the next attempt. When Implement or its independent build fails, Program-Agent also deletes the current build snapshot, resets the Workspace to the commit captured at the beginning of the round, and removes newly generated untracked files. Reusable Go caches are excluded from this cleanup. Build stdout and stderr from failed attempts are retained privately for diagnosis.

\paragraph{Last-stable recovery.}
Every independently built revision becomes the new last-stable state, which contains both its source commit and matching executable. On normal completion or after a later error or timeout, Program-Agent resets the Workspace to the last-stable source commit and restores its executable. If no implementation round ever passes independent build validation, any residual executable is removed so that an unvalidated intermediate candidate cannot be evaluated.





\section{Additional Experimental Details}\label{app:additional-experimental-details}

\subsection{Training Details}\label{app:training-details}
We fine-tune \texttt{Qwen3.8-27B} on trajectories collected from Program-Env for two epochs using ms-swift~\citep{zhao2024swiftascalablelightweightinfrastructure}. We issue task-level rollout requests for all 1,648 Program-Env tasks. Because Claude Code resets or switches its conversation context during a rollout, a single task-level rollout can yield multiple interaction segments with separate contexts. We treat each such segment as a separate training trajectory for supervised fine-tuning, yielding 64,505 training trajectories in total. Thus, the trajectory count does not correspond to 64,505 independent environment rollouts. All model parameters are updated during fine-tuning. Table~\ref{tab:training-config} summarizes the main training settings.

\begin{table}[htbp]
    \centering
    \small
    \caption{Main training settings for \texttt{Qwen3.8-27B}.
    The epoch count reflects the actual training duration.
    The warmup fraction follows the configured training schedule.}
    \label{tab:training-config}
    \begin{tabular}{ll}
        \toprule
        \textbf{Setting} & \textbf{Value} \\
        \midrule
        Base model & Qwen3.8-27B \\
        Training framework & ms-swift \\
        Training epochs & 2 \\
        Precision & BF16 \\
        Maximum sequence length & 262,144 tokens \\
        Sequence packing & Enabled (bin packing) \\
        Global batch size & 32 \\
        Micro-batch size & 1 \\
        Optimizer & Adam \\
        Learning rate & $1 \times 10^{-5}$ \\
        Minimum learning rate & $1 \times 10^{-6}$ \\
        Learning-rate schedule & Cosine decay \\
        Warmup fraction & 0.05 \\
        Weight decay & 0.1 \\
        Adam $\beta_1$, $\beta_2$ & 0.9, 0.95 \\
        Adam $\epsilon$ & $1 \times 10^{-8}$ \\
        Gradient clipping & 1.0 \\
        Random seed & 42 \\
        \bottomrule
    \end{tabular}
\end{table}

We pack training sequences into blocks of up to 262,144 tokens to accommodate long interaction trajectories. Examples exceeding the maximum sequence length are discarded. We compute the training loss only over eligible model-generated assistant tokens, while user inputs, environment feedback, and tool observations are used as conditioning context and excluded from the loss. Within assistant outputs, we additionally set the loss mask to zero for assistant turns that invoke tools unavailable in the corresponding environment. These invalid tool-use actions therefore do not contribute to the training objective, while the remaining valid turns from the same trajectory are retained. Training uses BF16 precision with full activation recomputation to reduce memory usage.

\subsection{Token Cost Analysis}\label{app:token-cost-analysis}
\begin{figure}[h] 
    \centering
    
    \includegraphics[width=.9\linewidth]{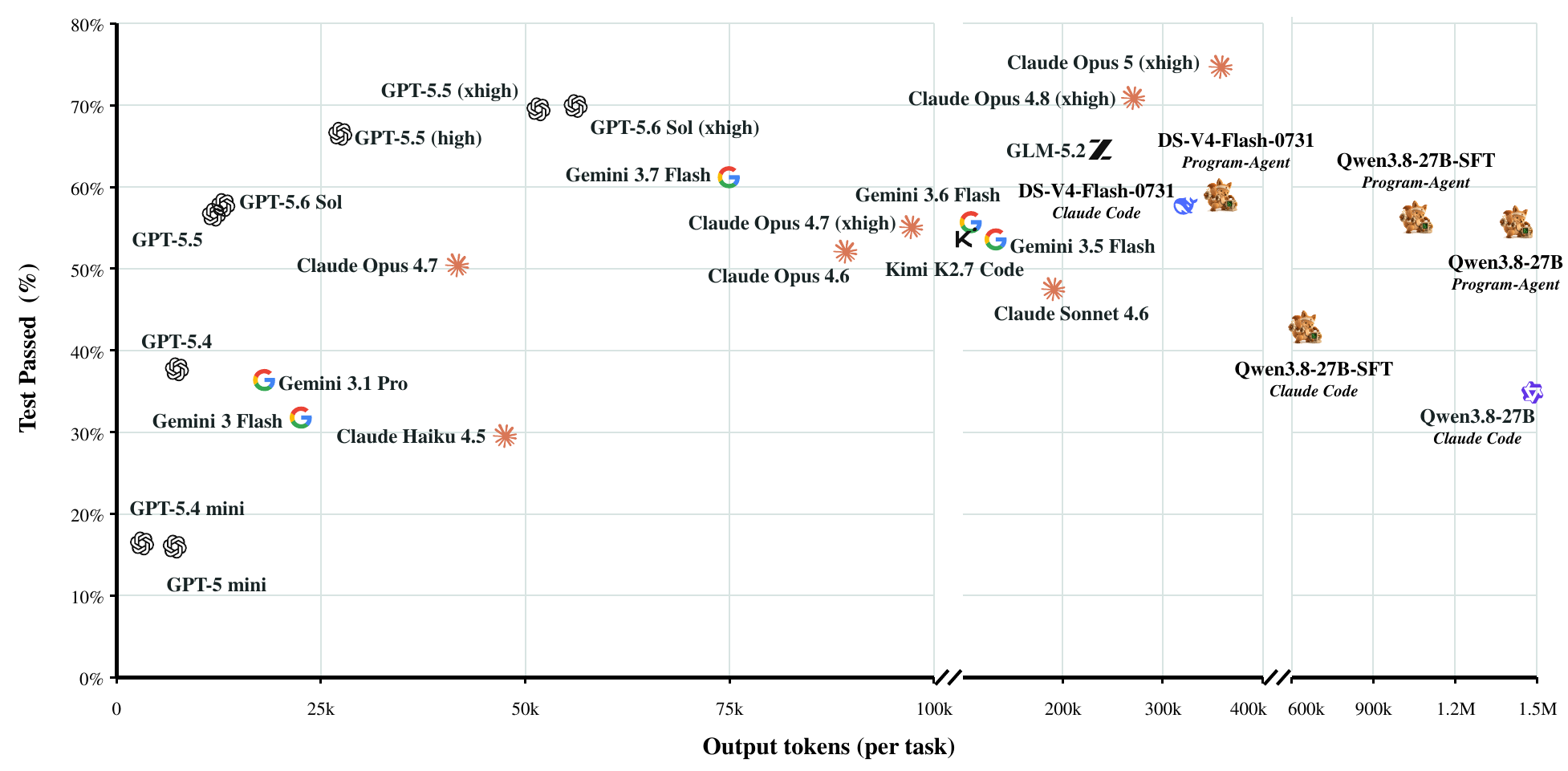} 
    \caption{Performance distribution of large language models on ProgramBench with respect to output token length and test pass rate.} 
    
    
    \label{fig:token_vs_score}
\end{figure}

\begin{figure}[h] 
    \centering
    
    \includegraphics[width=.9\linewidth]{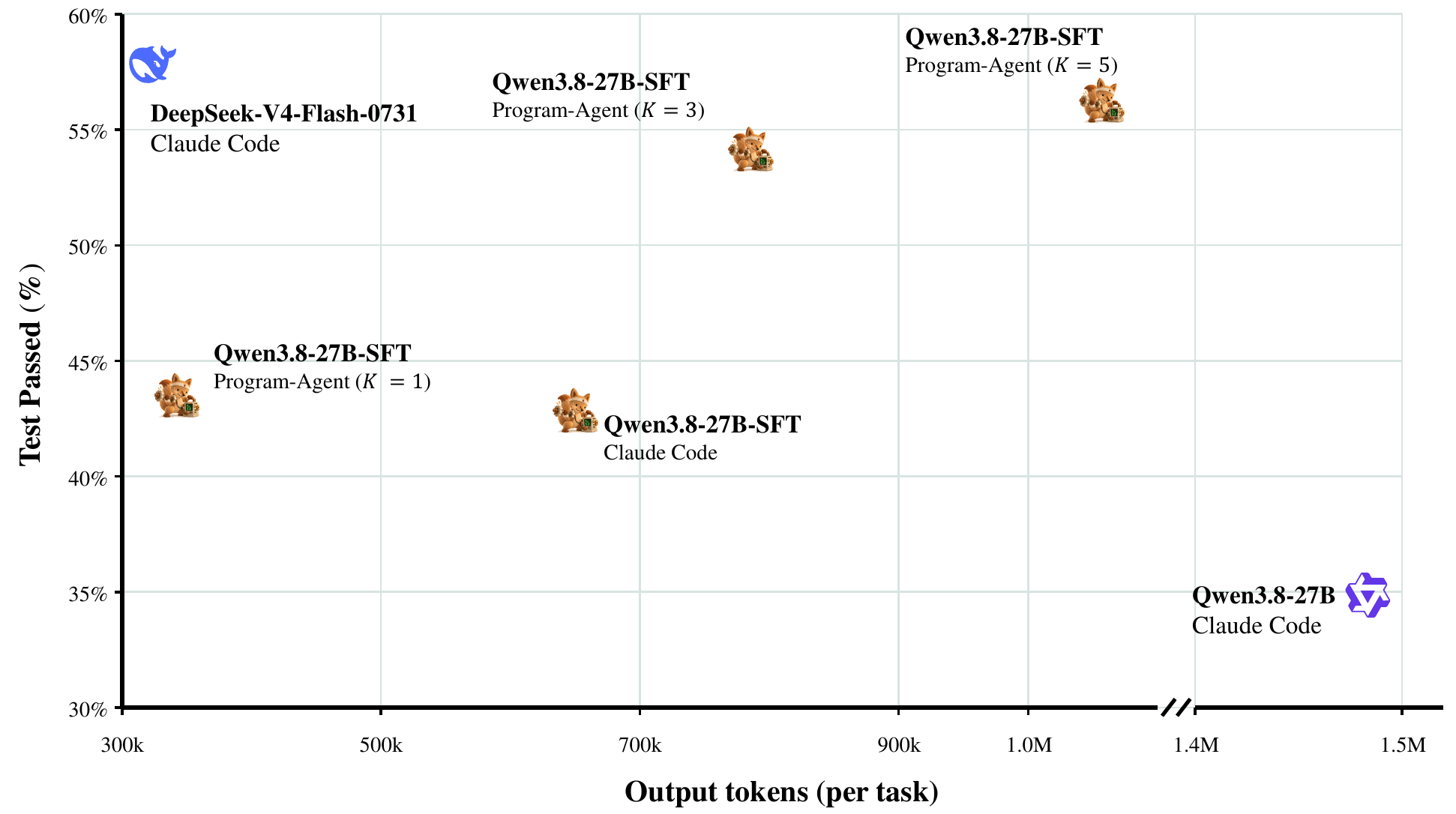} 
    \caption{Test pass rate versus output tokens across models and agents.
    } 
    
    
    \label{fig:token_itertaion}
\end{figure}
\paragraph{Program-Agent recovers more program behavior with fewer output tokens.}
Figure~\ref{fig:token_vs_score} compares Tests Passed against output tokens. Program-Agent raises \texttt{Qwen3.8-27B} from $34.87\%$ to $55.71\%$ and uses fewer tokens than \texttt{Claude Code}. It raises \texttt{Qwen3.8-27B-SFT} from $42.84\%$ to $56.25\%$. Both runs exceed flagship \texttt{Claude Opus 4.7} under \texttt{mini-SWE-agent}. Figure~\ref{fig:token_itertaion} shows that Program-Agent at $K{=}1$ already matches \texttt{Qwen3.8-27B-SFT} with \texttt{Claude Code} at a lower token cost. Extra rounds raise Tests Passed from $43.49\%$ at $K{=}1$ to $54.13\%$ at $K{=}3$ and $56.25\%$ at $K{=}5$. All three budgets still use fewer tokens than \texttt{Qwen3.8-27B} with \texttt{Claude Code}. These results indicate that Program-Agent spends tokens on structured revision rather than on a long single-session trajectory.

\subsection{Extra Analysis of Agent Behavior}\label{app:extra-analysis-of-agent-behavior}

\begin{figure}[h] 
    \centering

    \includegraphics[width=.6\linewidth]{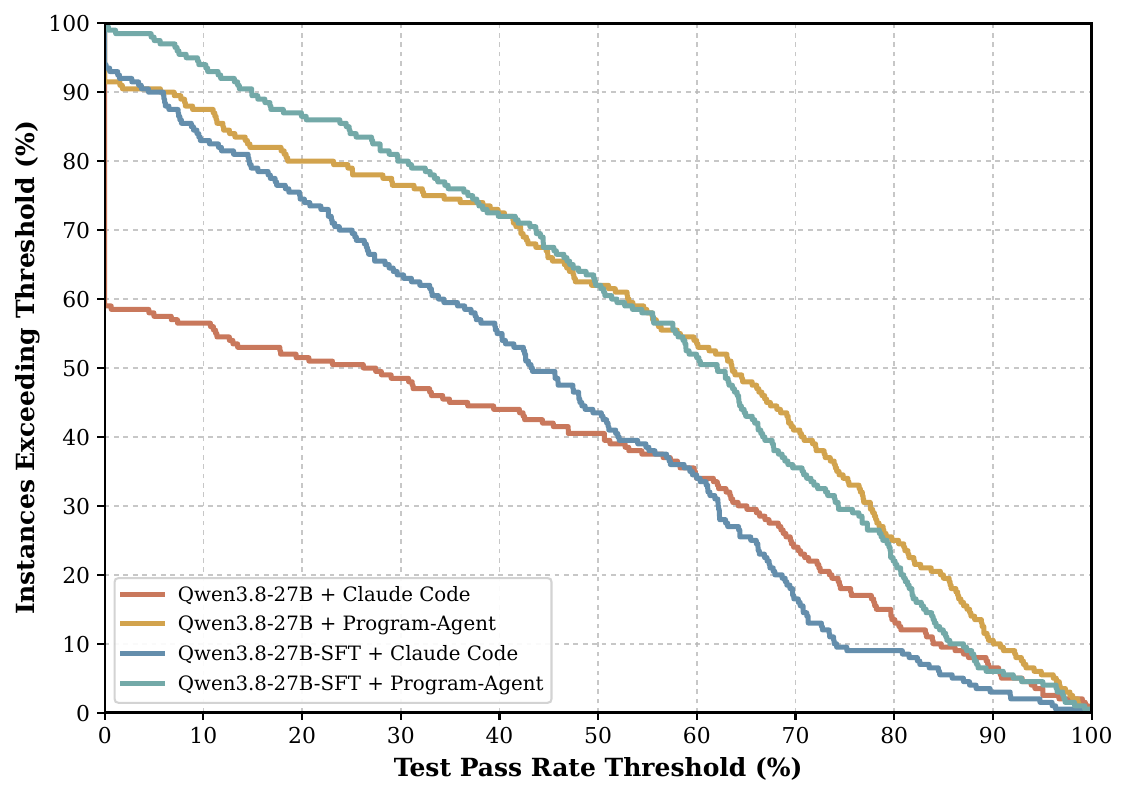}
    \caption{Cumulative distribution of test pass rates across models and agent scaffolds on ProgramBench.} 
    
    \label{fig:test_pass_cumulative_distribution}
\end{figure}

\paragraph{Structured iteration shifts the full distribution of solution quality.}
Figure~\ref{fig:test_pass_cumulative_distribution} shows that Program-Agent improves not just average results, but also how test pass rates are distributed across different tasks. When using \texttt{Qwen3.8-27B}, it increases the share of tasks that pass at least 50\% of tests from 40.5\% to 62.0\%, and raises the share passing at least 75\% of tests from 18.0\% to 34.0\%. With \texttt{Qwen3.8-27B-SFT}, the corresponding gains are from 43.5\% to 62.0\% and from 9.5\% to 29.5\%. The two Program-Agent curves stay close over most of the threshold range. This tells us its structured workflow with separate roles delivers more steady benefits, compared to simply switching between these two model versions. On the other hand, the effect of SFT changes depending on which test threshold we look at. SFT helps more tasks reach partial working functionality, but it cannot reliably boost the high-fidelity cases at the upper end of the distribution. The slow falling shape of all four curves also shows that tasks vary widely in difficulty. Meanwhile, the upward shift caused by Program-Agent demonstrates that this iterative development method works well across a wider range of task difficulty levels.

\begin{figure}[h] 
    \centering
    \includegraphics[width=.45\linewidth]{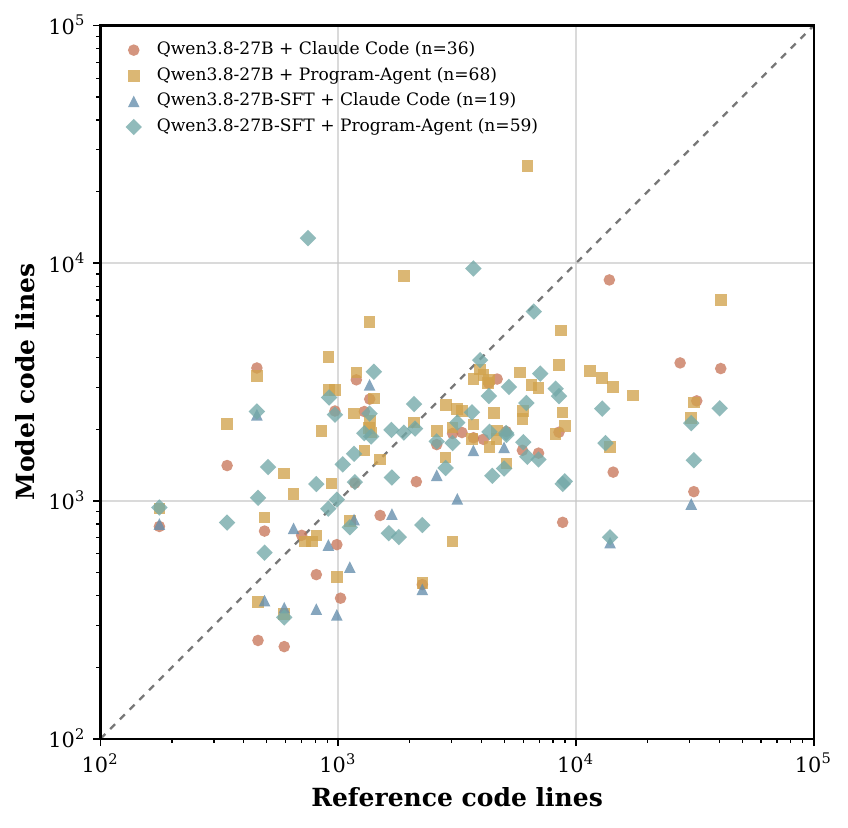}
    \caption{Reference vs. model code lines on a log–log scale. The \(y=x\) diagonal indicates parity. Only solutions passing $\geq$75\% of tests are shown.} 
    
    \label{fig:reference_vs_implementation_loc}
\end{figure}

\paragraph{Behavioral fidelity does not require reproducing the full reference codebase.}
To compare implementations that have roughly the same functions, we limit our code-size analysis to solutions that pass at least 75\% of the behavior tests. This gives us 182 test runs covering 93 tasks and four different model–agent setups. As shown in Figure~\ref{fig:reference_vs_implementation_loc}, these generated solutions have a median of 1,881 lines of code, while the matching reference implementations have 2,596 lines. In total, 67.6\% of the test runs sit below the \(y=x\) parity line, and the median ratio of model-generated code lines to reference code lines is 0.64. The implementations that exceed parity mostly come from smaller projects: their reference code has a median of 969 lines, while cases below parity have reference code with a median of 4,303 lines. This pattern tied to project size shows that small tasks are more impacted by fixed setup code and extra compatibility code. For large tasks, by contrast, agents can often recreate the required features just by coding the behavior visible through program interfaces and tests.

\begin{figure*}[h]
\centering
\begin{minipage}[t]{0.45\textwidth}
    \vspace{0pt}
    \centering
    \includegraphics[width=1.\linewidth]{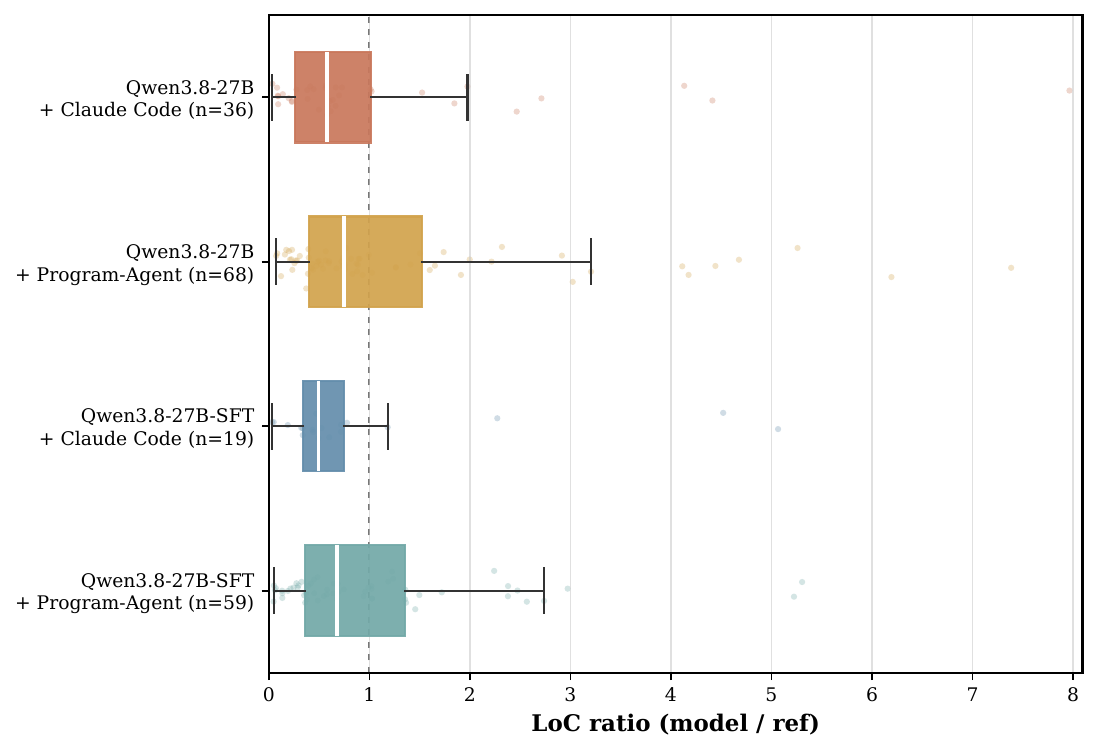}
    \captionof{figure}{
        LoC ratio (model / reference) by model–agent configuration for solutions passing $\geq$75\% of tests. The dashed line marks parity.
    }
    \label{fig:loc_ratio_by_configuration}
\end{minipage}
\hfill
\begin{minipage}[t]{0.45\textwidth}
    \vspace{0pt}
    \centering
    \includegraphics[width=1.0\linewidth]{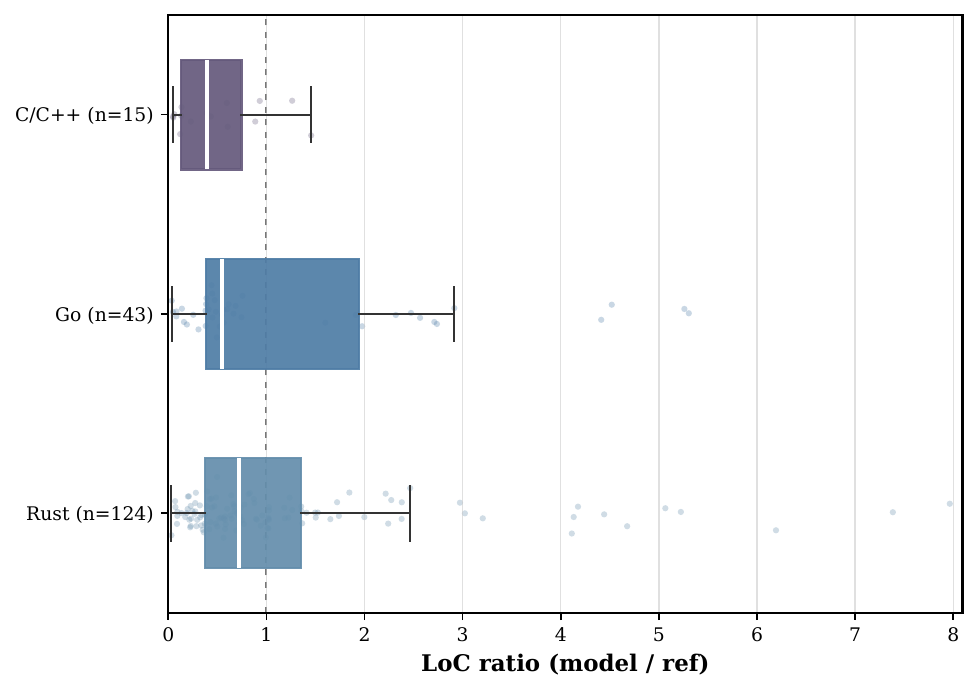}
    \captionof{figure}{
   LoC ratio (model / reference) by reference language for solutions passing $\geq$75\% of tests.
    }
    \label{fig:loc_ratio_by_reference_language}
\end{minipage}
\end{figure*}

\paragraph{Program-Agent trades compactness for more elaborate implementations.}
Figure~\ref{fig:loc_ratio_by_configuration} shows that the median code size for all four configurations stays below the parity level, but their distributions are quite different. When using \texttt{Qwen3.8-27B}, Program-Agent increases the median LoC ratio from 0.58 to 0.75, and when using \texttt{Qwen3.8-27B-SFT}, it raises the ratio from 0.49 to 0.67. Its upper quartile values also hit 1.53 and 1.36. By comparison, the matching \texttt{Claude Code} configurations only reach 1.01 and 0.75. This means Program-Agent achieves better functional performance not just by producing shorter code uniformly. Its iterative process often builds more complete and detailed implementations, and sometimes the generated code becomes larger than the reference version. When combined with the test-pass distributions shown in Figure~\ref{fig:test_pass_cumulative_distribution}, this finding indicates the extra code comes from repeated tries to handle behaviors that the first version of code fails to cover. Still, how large these final solutions end up varies a lot between different tasks.

\paragraph{Reference-language ecosystems shape reproduction compactness.}
Figure~\ref{fig:loc_ratio_by_reference_language} shows a clear trend that changes with programming language. C/C++ tasks have the lowest median ratio of 0.39, and only 13.3\% of the generated implementations meet or exceed the reference code size. Go falls in the middle with a median ratio of 0.55, while Rust is closest to the reference code size, with a median ratio of 0.72 and 35.5\% of generated solutions meeting or exceeding the reference size. One reasonable explanation is that agents can often recreate the functions of C/C++ utilities using more concise abstract code, without rebuilding all the low-level underlying code present in fully developed reference projects. On the other hand, Rust tasks are more commonly reimplemented using Rust code with a similar structure, so the new code keeps much of the verbosity seen in the original reference code. The long upper tails for Go and Rust in the figure also tell us that some tasks require large amounts of extra code to handle compatibility or embed behavioral data, which makes the rebuilt implementation far longer than the original reference code.


%% file: table/language_error.tex
\providecommand{\PBInlineCode}[1]{\texttt{\detokenize{#1}}}

\begin{table*}[t]
\centering
\caption{Selected examples of toolchain-version incompatibility.}
\label{tab:toolchain-mismatch}
\begingroup
\small
\setlength{\tabcolsep}{5pt}
\renewcommand{\arraystretch}{1.13}
\begin{tabular}{@{}
>{\centering\arraybackslash}m{.10\linewidth}
>{\raggedright\arraybackslash}m{.40\linewidth}
>{\raggedright\arraybackslash}m{\dimexpr.50\linewidth-4\tabcolsep\relax}
@{}}
\toprule
\textbf{Language}
& \textbf{Source of incompatibility}
& \textbf{Representative failure} \\
\midrule
\PBLanguageSmall{cpp}{C+}{C++}
& Build tools remove support for older policy versions.
& CMake~4.0 rejects \PBInlineCode{cmake_minimum_required}
with \PBInlineCode{VERSION 3.4} by default, because policy
compatibility below 3.5 is no longer supported. \\

\addlinespace[.35em]
\PBLanguageSmall{cpp}{C+}{C++}
& Standard-library availability depends on the selected
C++ language mode and library implementation.
& In C++17 mode, libc++ omits \PBInlineCode{std::auto_ptr}
and \PBInlineCode{std::random_shuffle} by default,
causing compilation failure. \\

\addlinespace[.35em]
\PBLanguageSmall{rust}{R}{Rust}
& Unstable features can change or be removed, making some
projects dependent on a dated nightly toolchain.
& Code enabling \PBInlineCode{box_syntax} fails with error
\PBInlineCode{E0557} on toolchains where that feature
has been removed. \\
\bottomrule
\end{tabular}
\endgroup
\end{table*}

%% file: table/compile_linter.tex
\providecommand{\PBInlineCode}[1]{\texttt{\detokenize{#1}}}
\begin{table*}[t]
\centering
\caption{Selected static checks on generated scripts. Preparation denotes \texttt{download.sh}; compilation denotes \texttt{compile.sh}.}
\label{tab:build-script-linter}
\begingroup
\small
\setlength{\tabcolsep}{5pt}
\renewcommand{\arraystretch}{1.10}
\begin{tabular}{@{}>{\raggedright\arraybackslash}p{.21\linewidth}>{\raggedright\arraybackslash}p{.12\linewidth}>{\raggedright\arraybackslash}p{\dimexpr.57\linewidth-6\tabcolsep\relax}>{\centering\arraybackslash}p{.10\linewidth}@{}}
\toprule
\textbf{Check} & \textbf{Scope} & \textbf{Pattern flagged} & \textbf{Level} \\
\midrule
Failure masking & Both
& Expressions that override a failed exit status, such as \PBInlineCode{|| true}. & Error \\
\addlinespace[.35em]
Error suppression & Both
& Redirections that discard standard error, such as \PBInlineCode{2>/dev/null}, except in simple tool-availability or version checks. & Error \\
\addlinespace[.35em]
Stream separation & Both
& Redirections that combine stdout and stderr, such as \PBInlineCode{2>&1} or \PBInlineCode{&>log}, except in simple tool-availability or version checks. & Error \\
\addlinespace[.35em]
Placeholder content & Both
& Standalone placeholder lines such as \PBInlineCode{...}, \PBInlineCode{<script>}, or \PBInlineCode{[truncated]}. & Error \\
\addlinespace[.35em]
Package installation & Compilation
& Package-manager installation, such as \PBInlineCode{apt}, \PBInlineCode{pip}, \PBInlineCode{npm}, or \PBInlineCode{cargo install <crate>}. Project install targets such as \PBInlineCode{make install} are allowed. & Error \\
\addlinespace[.35em]
Target artifact reference & Compilation
& The script omits a reference to the designated output artifact, named \PBInlineCode{executable}. & Error \\
\addlinespace[.35em]
Autotools bootstrap & Compilation
& \PBInlineCode{autoconf} with \PBInlineCode{./configure}, but no matched bootstrap command or auxiliary-file reference. & Error \\
\addlinespace[.35em]
Build commands & Preparation
& Configuration, compilation, or test commands, such as \PBInlineCode{cmake}, \PBInlineCode{make}, \PBInlineCode{cargo build}, and \PBInlineCode{go test}. & Error \\
\addlinespace[.35em]
Repository file access & Preparation
& Reads or existence checks matching repository dependency or build filenames, such as \PBInlineCode{Cargo.toml}, \PBInlineCode{go.mod}, and \PBInlineCode{Makefile}. & Error \\
\addlinespace[.35em]
Cargo manifest & Preparation
& \PBInlineCode{cargo fetch} without \PBInlineCode{--manifest-path} (Error), or with an explicit path not matched to a manifest generated by the script (Warning). & Error / Warning \\
\addlinespace[.35em]
Cargo fallback & Preparation
& \PBInlineCode{cargo fetch} followed through \PBInlineCode{||} by another fetch, creating an alternative prefetch path. & Error \\
\addlinespace[.35em]
\bottomrule
\end{tabular}
\endgroup
\end{table*}

%% file: table/heatmap-box.tex
\begin{figure*}[t]
\centering

\begin{minipage}[c]{0.53\textwidth}
    \vspace{0pt}
    \begin{tcolorbox}[
        cbox,
        title=Feedback template for executable generation,
        width=\linewidth,
        before skip=0pt,
        after skip=0pt,
        fontupper=\small
    ]
    \raggedright
    The previous attempt failed with status:
    \texttt{<status>}\\
    Failure reason: \texttt{<failure details>}\\[.4em]

    Required remediation for the next attempt:\\
    \texttt{<targeted repair instructions>}\\[.4em]

    Script linter markers:\\
    \texttt{- <rule> at <script>:<line>}\\
    \quad error: \texttt{<description>}\\
    \quad snippet: \texttt{<offending code>}\\
    \quad fix: \texttt{<suggested change>}\\[.4em]

    \texttt{<log excerpt>}
    \end{tcolorbox}
\end{minipage}
\hfill
\begin{minipage}[c]{0.44\textwidth}
    \centering
    \includegraphics[width=\linewidth]
        {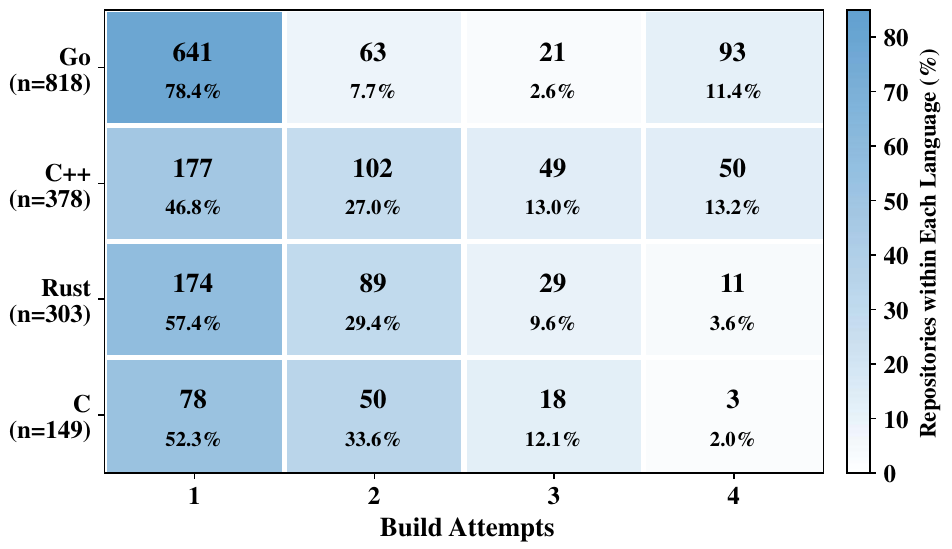}
\end{minipage}

\par\noindent
\begin{minipage}[t]{0.53\textwidth}
    \caption{Feedback template for executable generation.}
    \label{fig:build-feedback-template}
\end{minipage}
\hfill
\begin{minipage}[t]{0.44\textwidth}
    \caption{Build-attempt distribution across programming languages.}
    \label{fig:build-feedback-heatmap}
\end{minipage}

\end{figure*}

%% file: table/build_fix_suggestion.tex
\begin{longtable}{@{}>{\raggedright\arraybackslash}p{.28\linewidth}>{\raggedright\arraybackslash}p{\dimexpr.72\linewidth-2\tabcolsep\relax}@{}}
\caption{Suggested fixes attached to each finding of the linter check and the execution check.}

\label{tab:build-repair-guidance}\\
\toprule
\textbf{Finding} & \textbf{Suggested fix} \\
\midrule
\endfirsthead
\toprule
\textbf{Finding} & \textbf{Suggested fix} \\
\midrule
\endhead
\midrule
\multicolumn{2}{r@{}}{\footnotesize\textit{Continued on next page}} \\
\endfoot
\bottomrule
\endlastfoot
\addlinespace[.25em]
\multicolumn{2}{@{}l}{\textbf{Linter check}} \\
\addlinespace[.25em]
Failure masking
& Remove \texttt{|| true}. Handle commands that may legitimately fail with an explicit \texttt{if} or \texttt{case} branch. \\
\addlinespace[.35em]
Error suppression
& Remove \texttt{2>/dev/null}. Handle acceptable failures explicitly so that other errors remain visible. \\
\addlinespace[.35em]
Stream separation
& Keep standard output and standard error separate: leave them unredirected, or write them to separate logs (e.g., \texttt{>build.log 2>build.err}). \\
\addlinespace[.35em]
Placeholder content
& Replace the placeholder with the actual commands. \\
\addlinespace[.35em]
Package installation
& Move package-manager installation to \texttt{download.sh}. Keep only project build and install targets in \texttt{compile.sh}. \\
\addlinespace[.35em]
Target artifact reference
& Copy or link the built artifact to \texttt{./executable} at the end of \texttt{compile.sh}. \\
\addlinespace[.35em]
Autotools bootstrap
& Regenerate or restore \texttt{config.guess} and \texttt{config.sub} before \texttt{./configure}, for example via \texttt{autoreconf -fi}. \\
\addlinespace[.35em]
Build commands
& Move configuration, compilation, and test commands to \texttt{compile.sh}. Keep \texttt{download.sh} limited to installation, prefetching, and environment-variable exports. \\
\addlinespace[.35em]
Repository file access
& Embed the required manifest and lockfile contents in \texttt{download.sh}, so that preparation runs without access to the repository. \\
\addlinespace[.35em]
Cargo manifest
& Generate a self-contained \texttt{Cargo.toml} in \texttt{download.sh} and select it with \texttt{cargo fetch --manifest-path}. \\
\addlinespace[.35em]
Cargo fallback
& Replace the \texttt{||} chain with a single fetch against the generated manifest. \\
\addlinespace[.35em]
\addlinespace[.6em]
\multicolumn{2}{@{}l}{\textbf{Execution check: online preparation}} \\
\addlinespace[.25em]
Cargo cannot find \texttt{Cargo.toml}
& Generate a temporary manifest in \texttt{download.sh} and select it with \texttt{cargo fetch --manifest-path}. \\
\addlinespace[.35em]
Temporary manifest references missing targets
& Remove unnecessary target declarations or create the corresponding stub files. \\
\addlinespace[.35em]
Go cannot find \texttt{go.mod}
& Generate temporary module metadata and prefetch from that directory or with \texttt{-modfile}. \\
\addlinespace[.35em]
\addlinespace[.6em]
\multicolumn{2}{@{}l}{\textbf{Execution check: offline compilation}} \\
\addlinespace[.25em]
Missing headers, libraries, or build tools
& Install or prefetch the missing dependencies in \texttt{download.sh}, then rerun compilation. \\
\addlinespace[.35em]
Autotools cannot find \texttt{config.guess} or \texttt{config.sub}
& Restore the auxiliary files with \texttt{autoreconf -i} or the project's bootstrap script. \\
\addlinespace[.35em]
Go fails to obtain VCS information
& Add \texttt{-buildvcs=false} to the build command. \\
\addlinespace[.35em]
Compilation times out
& Check for dependency resolution or network waits; move the required preparation to \texttt{download.sh}. \\
\addlinespace[.6em]
\multicolumn{2}{@{}l}{\textbf{Execution check: artifact validation}} \\
\addlinespace[.25em]
Build completes without \texttt{./executable}
& Copy the selected target's artifact from its actual output path to \texttt{./executable}. \\
\addlinespace[.35em]
Generated artifact is not an ELF binary
& Export the selected target's native Linux executable. \\
\addlinespace[.35em]
\texttt{ldd} reports unresolved shared libraries
& Supply the missing runtime dependencies and required project libraries. \\
\end{longtable}

%% file: table/6type.tex
\begin{table}[h]
\centering
\caption{Interaction categories used in test generation. Each category directs a subagent to a specific behavioral aspect of the black-box executable.}
\label{tab:6type}
\begingroup
\small
\renewcommand{\arraystretch}{1.12}
\setbox0=\hbox{Subcommand}%
\edef\PBInteractionBodyWidth{\the\dimexpr\linewidth-\wd0-1.5em\relax}%
\begin{tabular}{@{}l@{\hspace{1.5em}}>{\raggedright\arraybackslash}p{\PBInteractionBodyWidth}@{}}
\toprule
\textbf{Type} & \textbf{Behavior to explore} \\
\midrule
Args
& Flag syntax, required and optional arguments, positional values, and argument type checks. \\
\addlinespace[.4em]
Config
& Settings read from environment variables and configuration files, including precedence when values conflict. \\
\addlinespace[.4em]
Help
& Output of \texttt{--help} and \texttt{-h}, usage summaries, and help for individual subcommands. \\
\addlinespace[.4em]
I/O
& Reading from standard input and files, separating standard output from standard error, exit status, and output formatting. \\
\addlinespace[.4em]
Subcommand
& Routing to the requested subcommand, resolving aliases, and applying global and subcommand-specific flags. \\
\addlinespace[.4em]
TUI
& Keyboard navigation, key bindings, screen updates, and transitions between interaction modes. \\
\bottomrule
\end{tabular}
\endgroup
\end{table}

%% file: table/table_en.tex
\begin{table*}[h]
\centering
\caption{Coverage instrumentation methods and cases beyond rule-based instrumentation.}
\label{tab:rule-instrumentation}
\begingroup
\small
\setlength{\tabcolsep}{4pt}
\renewcommand{\arraystretch}{1.12}
\begin{tabular}{@{}>{\centering\arraybackslash}m{.13\textwidth}>{\raggedright\arraybackslash}m{.40\textwidth}>{\raggedright\arraybackslash}m{\dimexpr.47\textwidth-4\tabcolsep\relax}@{}}
\toprule
\textbf{Language} & \textbf{Instrumentation method} & \textbf{Cases beyond rule-based instrumentation} \\
\midrule
\PBLanguage{c}{C}{C}{\PBGCCDoc}
& Add \PBCode{--coverage} to compilation and linking through
build variables, compiler wrappers, and recognized Makefiles or \PBCode{.mak} files.
Read \PBCode{.gcno/.gcda} with GNU \PBCode{gcov} or
\PBCode{llvm-cov gcov}, and summarize coverage with \PBCode{gcovr}.
& Build scripts may overwrite injected \PBCode{CFLAGS}/\PBCode{LDFLAGS}
or use custom \PBCode{.mak} variables that the language-specific rule does not reliably update.
Applying the same flags to every step can also break the build.
In kafel, for example, the injected \PBCode{--coverage} option is passed directly to
\PBCode{ld}, which does not accept it. \\
\addlinespace[.35em]
\PBLanguage{cpp}{C+}{C++}{\PBGCCDoc}
& Apply the same approach to C++ compilation and linking,
including \PBCode{CXXFLAGS} and compiler wrappers.
Support both GNU \PBCode{gcov} and \PBCode{llvm-cov gcov},
with \PBCode{gcovr} for reporting.
& Project build settings can overwrite injected flags, while direct compiler
calls can bypass the wrappers. Global injection can also interfere with build-system checks.
In fast-wfc, for example, a CMake compiler test is instrumented but linked without the
coverage runtime, causing configuration to fail before the project builds. \\
\addlinespace[.35em]
\PBLanguage{rust}{R}{Rust}{\PBRustDoc}
& Set \PBCode{RUSTFLAGS} to \PBCode{-C instrument-coverage}, then
merge \PBCode{.profraw} with \PBCode{llvm-profdata}
and export coverage with \PBCode{llvm-cov}.
& Replacing \PBCode{RUSTFLAGS} can remove conditional-compilation
or CPU-feature options required by the project, breaking its build,
so coverage flags must be merged with the existing options.
Targets without \PBCode{profiler_builtins} also need a toolchain
that provides the coverage runtime. \\
\addlinespace[.35em]
\PBLanguage{go}{Go}{Go}{\PBGoDoc}
& Build with \PBCode{-cover} and \PBCode{-coverpkg}.
Collect profiles under \PBCode{GOCOVERDIR} and convert them with
\PBCode{go tool covdata textfmt}.
& Matching a fixed set of command forms can miss calls such as
\PBCode{$GOROOT/bin/go}, so the build receives no coverage flags.
Building individual \PBCode{.go} files can also produce a package name
that differs from the repository import path, so the generated
\PBCode{-coverpkg} pattern selects no linked package. \\
\bottomrule
\end{tabular}
\par\vspace{.4em}
\begin{minipage}{\textwidth}
\footnotesize
\end{minipage}
\endgroup
\end{table*}

%% file: table/linter.tex
\begin{table*}[p]
\centering
\caption{Linter rules for rejecting weak oracles (1/2).}
\label{tab:linter-rule-inventory-1}
\begingroup
\small
\setlength{\tabcolsep}{4pt}
\renewcommand{\arraystretch}{1.12}
\begin{tabular}{@{}>{\raggedright\arraybackslash}p{.37\textwidth}>{\raggedright\arraybackslash}p{\dimexpr.53\textwidth-4\tabcolsep\relax}>{\centering\arraybackslash}p{.10\textwidth}@{}}
\toprule
\textbf{Rule} & \textbf{Detection pattern} & \textbf{Level} \\
\midrule
\addlinespace[.6em]
\multicolumn{3}{@{}l}{\textbf{Test structure}} \\
\addlinespace[.25em]
\texttt{test\_\allowbreak{}missing\_\allowbreak{}run\_\allowbreak{}sh} & Missing eval/run.sh. & Error \\
\texttt{test\_\allowbreak{}no\_\allowbreak{}discoverable\_\allowbreak{}pytest} & No test files, or a test file cannot be parsed. & Error \\
\texttt{no\_\allowbreak{}assertions} & No assert statement in the test function. & Error \\
\texttt{pass\_\allowbreak{}body} & The test body contains only pass statements. & Error \\
\texttt{trivially\_\allowbreak{}true} & A recognized always-true assertion. & Error \\
\addlinespace[.6em]
\multicolumn{3}{@{}l}{\textbf{Oracle strength}} \\
\addlinespace[.25em]
\texttt{sole\_\allowbreak{}returncode} & The sole assertion checks a zero exit code. & Error \\
\texttt{returncode\_\allowbreak{}in\_\allowbreak{}list} & An exit-code membership check against a list, tuple, or set. & Error \\
\texttt{all\_\allowbreak{}assertions\_\allowbreak{}weak} & All assertions use recognized weak signals: exit code, length, or isdigit. & Error \\
\texttt{generic\_\allowbreak{}keyword\_\allowbreak{}oracle} & A membership assertion uses a generic keyword such as success or error. & Error \\
\texttt{short\_\allowbreak{}substring} & A string literal in a membership assertion is shorter than 15 characters. & Warning \\
\texttt{weak\_\allowbreak{}sole\_\allowbreak{}assertion} & The sole assertion checks \texttt{len(x) > n} or \texttt{len(x) >= n} for an integer literal n. & Warning \\
\texttt{file\_\allowbreak{}exists\_\allowbreak{}no\_\allowbreak{}content} & File-existence assertion without a recognized content or size check. & Warning \\
\texttt{only\_\allowbreak{}negative\_\allowbreak{}assertions} & All assertions are negative checks. & Warning \\
\addlinespace[.6em]
\multicolumn{3}{@{}l}{\textbf{Control flow}} \\
\addlinespace[.25em]
\texttt{if\_\allowbreak{}else\_\allowbreak{}both\_\allowbreak{}assert} & Both sides of an if/else contain assertions. & Warning \\
\texttt{test\_\allowbreak{}swallowed\_\allowbreak{}failure} & The test launcher can mask pytest or coverage failure with \texttt{|| true}. & Error \\
\texttt{try\_\allowbreak{}except\_\allowbreak{}swallow} & An exception handler only passes, prints, or logs the exception. & Error \\
\texttt{assertion\_\allowbreak{}disjunction} & An assertion contains a Boolean or. & Warning \\
\texttt{if\_\allowbreak{}no\_\allowbreak{}else} & An if branch contains assertions but has no else branch. & Error \\
\texttt{for\_\allowbreak{}no\_\allowbreak{}guard} & An \texttt{all(...)} or negated \texttt{any(...)} assertion lacks a recognized preceding nonempty guard. & Warning \\
\texttt{any\_\allowbreak{}all\_\allowbreak{}no\_\allowbreak{}guard} & An any/all check lacks a recognized nonempty guard. & Warning \\
\texttt{xfail\_\allowbreak{}without\_\allowbreak{}strict\_\allowbreak{}reason} & An xfail marker lacks strict=True or a sufficiently specific reason. & Warning \\
\addlinespace[.6em]
\multicolumn{3}{@{}l}{\textbf{Golden integrity}} \\
\addlinespace[.25em]
\texttt{golden\_\allowbreak{}written\_\allowbreak{}in\_\allowbreak{}test} & A test mentioning golden data contains a recognized file-write operation. & Error \\
\texttt{golden\_\allowbreak{}no\_\allowbreak{}equality} & Golden data is mentioned without a recognized equality comparison. & Error \\
\texttt{golden\_\allowbreak{}docstring} & The docstring mentions golden data that is absent from the test body. & Error \\
\addlinespace[.6em]
\multicolumn{3}{@{}l}{\textbf{Runtime stability}} \\
\addlinespace[.25em]
\texttt{test\_\allowbreak{}network\_\allowbreak{}dependency} & The test contains recognized networking references. & Error \\
\texttt{test\_\allowbreak{}time\_\allowbreak{}dependency} & The test references time.time, datetime.now, or date.today. & Error \\
\texttt{test\_\allowbreak{}random\_\allowbreak{}without\_\allowbreak{}seed} & The test uses random without an explicit random.seed. & Error \\
\addlinespace[.6em]
\multicolumn{3}{@{}l}{\textbf{Fixture packaging}} \\
\addlinespace[.25em]
\texttt{skip\_\allowbreak{}without\_\allowbreak{}fixture} & A test skips when fixture assets are missing. & Error \\
\texttt{missing\_\allowbreak{}fixture\_\allowbreak{}assets} & A recognized fixture path does not exist in the delivered branch. & Error \\
\bottomrule
\end{tabular}
\endgroup
\end{table*}
\clearpage
\makeatletter
\providecommand{\PBLintContinuationTop}[1]{%
  \if@twocolumn
    \twocolumn[{#1}]%
  \else
    \par\noindent #1\par
  \fi
}
\PBLintContinuationTop{%
\begin{minipage}{\textwidth}
\def\@captype{table}
\centering
\caption{Linter rules for rejecting weak oracles (2/2).}
\label{tab:linter-rule-inventory-2}
\begingroup
\small
\setlength{\tabcolsep}{4pt}
\renewcommand{\arraystretch}{1.12}
\begin{tabular}{@{}>{\raggedright\arraybackslash}p{.37\textwidth}>{\raggedright\arraybackslash}p{\dimexpr.53\textwidth-4\tabcolsep\relax}>{\centering\arraybackslash}p{.10\textwidth}@{}}
\toprule
\textbf{Rule} & \textbf{Detection pattern} & \textbf{Level} \\
\midrule
\addlinespace[.6em]
\multicolumn{3}{@{}l}{\textbf{Black-box access}} \\
\addlinespace[.25em]
\texttt{test\_\allowbreak{}imports\_\allowbreak{}source\_\allowbreak{}module} & The test file imports source or a source submodule. & Error \\
\texttt{test\_\allowbreak{}runtime\_\allowbreak{}source\_\allowbreak{}tree\_\allowbreak{}dependency} & The launcher or test references /workspace/source at runtime. & Error \\
\texttt{test\_\allowbreak{}executable\_\allowbreak{}path\_\allowbreak{}contract} & The test uses a recognized fragile relative executable path. & Error \\
\addlinespace[.6em]
\multicolumn{3}{@{}l}{\textbf{Output comparison}} \\
\addlinespace[.25em]
\texttt{float\_\allowbreak{}exact\_\allowbreak{}match} & A pattern suggests exact comparison of floating-point output. & Warning \\
\texttt{unordered\_\allowbreak{}output\_\allowbreak{}assumption} & A pattern suggests an ordering assumption for map-like output. & Warning \\
\texttt{rendering\_\allowbreak{}exact\_\allowbreak{}match} & The test contains image, pixel, screenshot, or hash-related patterns. & Warning \\
\addlinespace[.6em]
\multicolumn{3}{@{}l}{\textbf{Test organization}} \\
\addlinespace[.25em]
\texttt{coverage\_\allowbreak{}chasing\_\allowbreak{}artifact} & Names or test text match coverage-chasing or round-artifact patterns. & Warning \\
\texttt{catches} & The CATCHES description is absent or shorter than 15 characters. & Warning \\
\bottomrule
\end{tabular}
\endgroup
\end{minipage}
\par\vspace{1em}
}
\makeatother

%% file: table/backends.tex
\begin{table*}[t]
\centering
\caption{Language-specific backends for computing core coverage.}
\label{tab:test-feedback-backends}
\begingroup
\small
\setlength{\tabcolsep}{5pt}
\renewcommand{\arraystretch}{1.13}
\begin{tabular}{@{}>{\raggedright\arraybackslash}p{.12\linewidth}>{\raggedright\arraybackslash}p{.62\linewidth}>{\raggedright\arraybackslash}p{\dimexpr.26\linewidth-4\tabcolsep\relax}@{}}
\toprule
\textbf{Language} & \textbf{Collection and export} & \textbf{Counting unit} \\
\midrule
C / C++
& GNU \texttt{gcov} or \texttt{llvm-cov gcov} reads \texttt{.gcno}/\texttt{.gcda}; \texttt{gcovr} exports coverage reports.
& Executable source lines. \\
\addlinespace[.4em]
Rust
& LLVM source-based coverage; \texttt{llvm-profdata} merges \texttt{.profraw}, and \texttt{llvm-cov} exports line counts in LCOV format.
& Executable source lines. \\
\addlinespace[.4em]
Go
& Runtime counters in \texttt{GOCOVERDIR} are converted by \texttt{go tool covdata textfmt} to a coverage profile.
& Basic blocks weighted by statement count. \\
\bottomrule
\end{tabular}
\par\vspace{.35em}
\begin{minipage}{\linewidth}
\footnotesize
\textit{Note.} Backends share a feedback schema while retaining their counting semantics. For Rust, line counts are derived from LLVM’s source-region coverage mappings.
Go statement coverage is not relabeled as line coverage. Source locations support target navigation. 
\end{minipage}
\endgroup
\end{table*}

%% file: table/summary.tex
\newtcblisting{templatebox}[2]{
 cbox,
  listing only,
  title=#1,
  label=#2,
  listing options={
    basicstyle=\ttfamily\footnotesize,
    breaklines=true,
    columns=fullflexible,
    keepspaces=true,
  },
}

\begin{templatebox}{Template of \texttt{summary.md} written by each subagent}{lst:summary-template}
# Eval Summary

## Assignment
- Role / Module / Assigned source paths / Iteration

## Validation Repairs
- Repair status
- Repaired tests
- Remaining repair items

## Coverage Target Plan
| Target | Root entry | Input/state | Oracle | Test | Result |
| --- | --- | --- | --- | --- | --- |
For every skipped target, record the control condition, probe input, observed result, and decision.

## Generated Artifacts
- eval/run.sh
- Test files / Runtime resources / Golden outputs

## Verification
- Golden self-check command and result
- eval/results.xml

## Remaining Limitations
- None, or concrete unresolved reachability or stability issues
\end{templatebox}

%% file: table/feedback_file_tree.tex
\begin{tcolorbox}[cbox,title={Test feedback directory structure},
]
\begingroup
\small
\begin{verbatim}
|-- Feedback_Docs/
|   |-- modules/
|   |   `-- module_XX/
|   |       |-- validation.json
|   |       |-- module_assignment.json
|   |       |-- coverage_targets.json
|   |       |-- coverage_summary.md
|   |       |-- summary.md
|   |       |-- static_analysis_result/        
|   |       |   |-- status.json
|   |       |   |-- summary.json
|   |       |   |-- functions/*.json
|   |       |   |-- callgraph/*.json
|   |       |   |-- controlflow/*.json
|   |       |   `-- dataflow/*.json
|   `-- upstream_harvest/
|       |-- round_context.json
|       |-- validation.json
|       |-- module_assignment.json
|       |-- remaining_upstream_tests.json
|       `-- summary.md
\end{verbatim}
\endgroup
\end{tcolorbox}

%% file: table/early_stop.tex
\begin{table}[H]
\centering
\caption{Termination settings for test generation.}
\label{tab:test-generation-stopping}
\small
\renewcommand{\arraystretch}{1.15}
\begin{tabular}{@{}l@{\hspace{2em}}l@{}}
\toprule
\textbf{Parameter} & \textbf{Setting} \\
\midrule
Functionality coverage threshold & 80\% \\
Iteration limit & 5--20 rounds\\
Earliest early-stopping round & Round 5 \\
Early-stopping window & 3 rounds \\
\bottomrule
\end{tabular}
\end{table}

%% file: table/8type.tex
\begin{table}[h]
\centering
\small
\renewcommand{\arraystretch}{1.1}
\caption{Functional categories of the 1,648 Program-Env repositories.}
\begin{tabular}{@{}>{\raggedright\arraybackslash}p{0.24\linewidth}>{\raggedright\arraybackslash}p{0.64\linewidth}r@{}}
\toprule
\textbf{Category} & \textbf{Description} & \textbf{Repos} \\
\midrule
Developer Tools & Build, test, lint and format, version control, documentation, web frameworks, and general-purpose libraries. & 441 \\
\addlinespace[0.4em]
CLI \& System Utilities & File and disk management, process and device control, container and cluster management, networking, text processing, compression, and terminal tools. & 372 \\
\addlinespace[0.4em]
Data \& Scientific Computing & Databases, structured data processing, numerical computation, simulation, bioinformatics, machine learning, and NLP. & 243 \\
\addlinespace[0.4em]
Security \& Forensics & Vulnerability scanning, authentication, cryptography and key management, auditing, reverse engineering, and digital forensics. & 227 \\
\addlinespace[0.4em]
Media \& Graphics & Image, audio, and video processing, 3D rendering, plotting, and GUI or TUI components. & 115 \\
\addlinespace[0.4em]
Games \& Education & Games and game engines, programming tutorials, exercises, terminal animations, and demos. & 110 \\
\addlinespace[0.4em]
Productivity \& Communication & Notes and to-do management, readers and chat, clipboard tools, publishing, and end-user office applications. & 73 \\
\addlinespace[0.4em]
Languages \& Runtimes & Language implementations, parsers, virtual machines, REPLs, language servers, debuggers, and runtime managers. & 67 \\
\bottomrule
\end{tabular}
\label{tab:functional-categories}
\end{table}

%% file: table/pb-agent_pseudo-code.tex
\begin{algorithm}[!htbp]
\caption{Normal phase orchestration of Program-Agent}
\label{alg:program-agent-orchestration}
\begin{algorithmic}[1]

\Require Task information $D$, reference program $B^\star$,
         implementation budget $K$

\Ensure Build-validated source revision $C_K$, including \texttt{compile.sh}

\State $A \gets \textsc{Analysis}(D,B^\star)$

\State Freeze $A$ as a read-only Handoff artifact

\State Hide $B^\star$ from implementation-visible paths

\For{$k \gets 1$ to $K$}

    \State $(C_k,I_k) \gets
    \textsc{Implement}(D,A,R_1,\ldots,R_{k-1},W)$

    \State $E_k \gets \textsc{IndependentBuild}(C_k)$

    \State Attach the build record for $(C_k,E_k)$ to $I_k$
    \State Freeze $I_k$ as a read-only Handoff artifact

    \If{$k<K$}

        \State $R_k \gets
        \textsc{Review}(D,A,R_1,\ldots,R_{k-1},I_k,B^\star,E_k)$

        \State Freeze $R_k$ as a read-only Handoff artifact
    \EndIf
\EndFor

\State Restore $W$ to source revision $C_K$

\State \Return the source tree at $C_K$, including \texttt{compile.sh}

\end{algorithmic}
\end{algorithm}

%% file: table/pb-agent_layout.tex
\begin{tcolorbox}[
  cbox,
  breakable,
  title={Logical artifact layout of Program-Agent}
]
\begin{lstlisting}[
  basicstyle=\small\ttfamily,
  columns=fullflexible,
  breaklines=true,
  breakatwhitespace=false,
  breakindent=0pt,
  breakautoindent=false,
  showspaces=false,
  showstringspaces=false,
  aboveskip=0pt,
  belowskip=0pt
]
/workspace/
|-- src/
|-- build/
|-- compile.sh
`-- executable                 # derived build artifact

/handoff/
|-- analysis-initial/
|   |-- analysis.md
|   `-- ...
|-- implementation-001/
|   |-- delivery.md
|   |-- independent-build.json
|   `-- ...
|-- review-001/
|   |-- review.md
|   `-- ...
`-- implementation-002/
    `-- ...

/var/lib/harbor-agent/workflow/ # private during execution
|-- builds/
|   `-- rev-001/
|       |-- source/
|       |-- build.stdout
|       |-- build.stderr
|       `-- build.json
|-- failures/
|-- sessions/
|-- streams/
|-- phases.json
|-- state.json
`-- last-stable.json
\end{lstlisting}
\end{tcolorbox}

%% file: table/pb-agent_access.tex
\begin{table*}[!htbp]
\centering
\small
\caption{Resource access across Program-Agent stages.}
\label{tab:program-agent-access}
\setlength{\tabcolsep}{6pt}
\renewcommand{\arraystretch}{1.12}

\begin{tabularx}{\textwidth}{
    @{}
    >{\raggedright\arraybackslash}p{0.22\textwidth}
    >{\raggedright\arraybackslash}X
    >{\raggedright\arraybackslash}X
    >{\raggedright\arraybackslash}X
    @{}
}
\toprule
Resource & Analysis & Implement & Review \\
\midrule

\multicolumn{4}{@{}l}{\textit{Executable access}} \\
\addlinespace[0.15em]

Reference executable
&
Execute only
&
Unavailable
&
Execute only
\\

Candidate executable
&
Unavailable
&
Working candidate
&
Immutable copy
\\

\addlinespace[0.5em]
\multicolumn{4}{@{}l}{\textit{Source workspace}} \\
\addlinespace[0.15em]

Product source
&
Disposable copy
&
Shared Workspace; read/write
&
Disposable copy
\\

\addlinespace[0.5em]
\multicolumn{4}{@{}l}{\textit{Handoff artifacts}} \\
\addlinespace[0.15em]

Active stage
&
Read/write
&
Read/write
&
Read/write
\\

Completed stages
&
Not applicable
&
Read-only
&
Read-only
\\

\addlinespace[0.5em]
\multicolumn{4}{@{}l}{\textit{Private orchestration state}} \\
\addlinespace[0.15em]

Workflow state
&
Inaccessible
&
Inaccessible
&
Inaccessible
\\

\bottomrule
\end{tabularx}
\end{table*}

%% file: table/pb-agent_validation.tex
\begin{table}[!htbp]
\centering
\small
\caption{Machine-checked stage completion conditions.}
\label{tab:program-agent-validation}
\setlength{\tabcolsep}{5pt}
\renewcommand{\arraystretch}{1.12}

\begin{tabularx}{\linewidth}{
    @{}
    >{\raggedright\arraybackslash}p{0.18\linewidth}
    >{\raggedright\arraybackslash}X
    @{}
}
\toprule
Stage & Completion conditions \\
\midrule

Analysis
&
\texttt{analysis.md} exists and is non-empty.
\\

Implement
&
\texttt{delivery.md} exists and is non-empty; tracked files contain no staged
or unstaged changes; \texttt{compile.sh} is committed; and a clean source
snapshot independently produces an executable.
\\

Review
&
\texttt{review.md} exists and is non-empty.
\\

\bottomrule
\end{tabularx}
\end{table}